\documentclass[aps]{revtex4}
\pdfoutput=1
\usepackage{eurosym}
\usepackage{amsfonts}
\usepackage{amsmath}
\usepackage{amssymb,epsf}
\usepackage{color}
\usepackage{graphicx}
\usepackage{epstopdf}
\usepackage{float}
\usepackage{caption}
\usepackage{subfig}
\usepackage{hyperref}

\makeatletter
\newbox\sf@box
{\def\sf@one{#1}%
	\def\sf@two{#2}%
	\setbox\sf@box\hbox
	\bgroup}%
{ \egroup
	\ifx\@empty\sf@two\@empty\relax
	\def\sf@two{\@empty}
	\fi
	\ifx\@empty\sf@one\@empty\relax
	\subfloat[\sf@two]{\box\sf@box}%
	\else
	\subfloat[\sf@one][\sf@two]{\box\sf@box}%
	\fi}
\makeatother

\makeatletter
\renewcommand{\p@subfigure}{\thefigure-}
\makeatother

\definecolor{aogreen}{rgb}{0.0, 0.5, 0.0}

\def\ketm#1{  \left\vert  #1   \right\rangle   }

\def\bram#1{  \left\langle  #1   \right\vert   }

\begin{document}

\title{Controlling quantum random walk with a step-dependent coin}
\author{S. Panahiyan$^{1}$ \footnote{%
email address: shahram.panahiyan@uni-jena.de} and S. Fritzsche$^{1,2}$%
\footnote{%
email address: s.fritzsche@gsi.de} }

\affiliation{$^1$Helmholtz-Institut Jena, Fr\"{o}belstieg 3, D-07743 Jena, Germany  \\
             $^2$Theoretisch-Physikalisches Institut, Friedrich-Schiller-University Jena, D-07743 Jena, Germany}

\begin{abstract}
We report on the possibility of controlling quantum random walks with a step-dependent coin. The coin is characterized by a (single) rotation angle. Considering different rotation angles, one can find diverse probability distributions for this walk including: complete localization, Gaussian and asymmetric likes. In addition, we explore the entropy of walk in two contexts; for probability density distributions over position space and walker's internal degrees of freedom space (coin space). We show that entropy of position space can decrease for a step-dependent coin with the step-number, quite in contrast to a
walk with step-independent coin. For entropy of coin space, a damped oscillation is found for walk with step-independent coin while for a step-dependent coin case, the behavior of entropy depends on rotation angle. In general, we demonstrate that quantum walks with simple initiatives may exhibit a quite complex and varying behavior if step-dependent coins are applied. This provides the possibility of controlling quantum random walk with a step-dependent coin.  
\end{abstract}

\maketitle

\section{Introduction}

Classical random walks (CW) are known (stochastic) processes that
have been used for developing stochastic algorithms
\cite{Motwani}, in studying biological behavior \cite{Berg}, for
image segmentation \cite{Grady} as well as in earthquake
simulations \cite{Helmstetter}. A classical random walk is
typically described in terms of a coin and some particular rule,
how the outcome of a coin toss affects the motion of a walker. For
different possible outcomes of the coin, the walker will then move
to the sites available for it. In general, a Gaussian probability
density distribution is obtained for most of the CWs.

In a quantum random walk (QW), in contrast, the behavior of the
walker is governed by quantum mechanics \cite{Aharonov}. Both, the
coin and the movement (shift) of the walker are represented by two
operators that are applied to the walker at each step of the walk:
While the shift operator moves the walker simply to the left or
right on a mesh, the (so-called) coin operator just acts on the
internal degrees of freedom of the walker. In a QW, therefore,
the walker directly interacts with the coin and this leads to a
number of properties such as: a ballistic spread of the walker's
wave function or a non-Gaussian probability density distribution
\cite{Aharonov}. In fact, QWs have been found to provide an
efficient framework for: developing better (quantum) algorithms
\cite{Ambainis2,Kendon2}, increasing the processing power for
solving problems \cite{Ambainis3} and in simulating complex
physical systems \cite{Mohseni}. It was shown, moreover, that QW
can be utilized as a universal computational primitives
\cite{Childs,Lovett} in order to simulate other quantum processes.
In addition, various experimental realization of QWs were
demonstrated by means of ultracold atoms \cite{Karski}, photons
\cite{Schreiber,Broome} or ions\cite{Schmitz,Zahringer} could
simulate a nontrivial one-dimensional topological phase
\cite{Kitagawa,Obuse,Rakovszky}.

Because of these properties, different classes of QWs have been
explored in the past decade, including walks with multiple
\cite{Brun3} or decoherent coins \cite{Brun1}, aperiodic walks
\cite{Ribeiro}, walks in two and more dimensions \cite{Mackay} or
with multiple walkers \cite{Pathak}. Moreover, the concept of
QWs have been applied to position-dependent walks
\cite{Suzuki}, to time-dependent walks that remember their history
\cite{Flitney,Banuls,Xue,Montero2} as well as to (so-called) spinor
Bose-Einstein condensate \cite{Alberti2,Alberti}.

Since QW can be utilized for simulating other quantum systems or
to program and engineer different quantum algorithms
\cite{Kendon,Venegas-Andraca}, a high control is desirable on the
behavior and time-evolution of the walker. In order to enhance it,
\textit{decoherent} walks have been proposed in the literature
\cite{Kendon,Venegas-Andraca} i.e. walks with a (non-unitary) coin
which introduces some coupling with the walker's environment.
However, such a coupling requires the use of density operators and
makes the computations very complex. Typically, such a decoherence
does not enable one (so easily) to further explore the symmetry of
the system, and may result into a loss of (parts of the)
probability density and unitarity nature of process
\cite{Kendon,Venegas-Andraca}. In addition, the reliability,
complexity and efficiency in simulating quantum systems with
decoherence have been challenged, cf.~Ref. \cite{Georgescu}.
Therefore, the question remains to which extent, we can have a
(coherent) walk with just unitary coin and shift operators and for
which the desired probability density distribution is obtained by
adjusting one (or a few) parameters. Below, we demonstrate that
such a control and behavior of QW can be achieved by applying a
\textit{step-dependent coin}.

Another question of this work refers to the amount of information
that can be processed by a QW. Since every (quantum) process that
transforms the state of a system also changes the amount of
information, it is important to understand how a step-dependent
coin affects the information of the system, compared to an usual
QW. To address this question, we shall study and discuss below
also the Shannon entropy of the walks \cite{Nielsen}. The Shannon
entropy was first introduced in 1948 by Claude Shannon
\cite{Shannon} in order to measures the amount of uncertainty that
is present in the state of a physical system. Here, we show that
walks with a step dependent coin may exhibit a diverse behavior of
the entropy, quite different from the usual QW with
step-independent coins.

The outline of the paper is as follows. In the next sections, we
first introduce our walk with a step-dependent coin and recall
some of its properties. In Section \ref{Simulation}, then, simulations are
performed and analyzed for different coins, together with a short
classification of the corresponding probability distributions. We
also discuss a possible experimental realization of such a QW in
section \ref{Experimental} and the entropy as a function of the number of steps in
section \ref{Shannon Entropy}. A brief summary and conclusions are finally given in
section \ref{Conclusion}.
\section{Quantum random walk with step-dependent coin}     \label{Quantum}

When compared with CW, many features of QW arise from the
interplay of the coin and walker, i.e.\ from the walker's
\textit{internal} degrees of freedom. Due to this interplay, QW
may exhibit a much richer behavior of its probability distribution
than CW (if different coins and initial states are considered).
Of course, this freedom in choosing the coin and state of the
walker can be employed also to find novel probability density
distributions for the walker after a given number of steps which
is one of central motivations of this investigation. We here apply
a \textit{step-dependent coin} of the form
\begin{eqnarray}
   \widehat{C} & = & \cos T\theta\: \ketm{0}_{C} \bram{0} \:+\:
                     \sin T\theta\: \ketm{0}_{C} \bram{1}                              \notag \\[0.1cm]
\label{coin}
               &   & \quad +\:
                     \sin T\theta\: \ketm{1}_{C} \bram{0} \:-\:
                     \cos T\theta\: \ketm{1}_{C} \bram{1} \: ,
\end{eqnarray}
in which $T$ refers to the number of step and where the rotation
angle $\theta$ is characteristic for the given walk. The factor 
which enables controlling the walk is $T\theta$. Such a
step-dependent coin can be realized, for example, with photons if
their polarization are rotated by the (step-dependent) angle
$T\theta$ \cite{Schreiber,Broome}. Such a step-dependent
'rotation' can be achieved also with trapped ions if proper
(resonant) radio-frequency pulses are applied to the internal
state of the ions \cite{Schmitz}. Below, we shall therefore refer
to $\theta$ simply as the (fixed) \textit{rotation} angle that is
characteristic for a particular coin. The coin
operator (\ref{coin}) is unitary and independent of the particular
choice of $\theta$ and number of steps, and thus reversible
\cite{Zahringer}. Indeed, this property helps simplify many
computations and ensures the symmetries which are required for
exploring topological phases \cite{Kitagawa} and for employing it
as a \textit{single-qubit} gate \cite{Nielsen}.

For a (single-qubit) walker on a linear mesh, the Hilbert space,
$\mathcal{H}_{C}$ of the \textit{coin} space is simply spanned by
$\{ \ketm{0},\: \ketm{1} \}$. The overall operator of the
(quantum) walk consists out of two parts: i) the \textit{coin}
operator that just acts upon the internal degrees of freedom and
ii) the \textit{shift} operator that moves the walker along the
mesh, in dependence on the internal state of the walker. Here, we
consider the (standard) conditional shift operator
\begin{eqnarray}
   \widehat{S} & = & \ketm{0}_{C} \bram{0} \otimes \sum \ketm{i+1}_{P} \bram{i}            \notag \\[0.1cm]
               &   & \quad +\:
                     \ketm{1}_{C} \bram{1} \otimes \sum \ketm{i-1}_{P} \bram{i}\, ,
\end{eqnarray}
and where the position (Hilbert) space, $\mathcal{H}_{P}$, is
spanned by $\{ \ketm{i}_{P}: i\in \mathbb{Z}\}$. Of course, the
QW proceeds overall within the product space $\mathcal{H}
\:\equiv\: \mathcal{H}_{P} \otimes \mathcal{H}_{C}$ and by
applying successively the evolution operator $\widehat{U} \:=\:
\widehat{S}\; \widehat{C}$ upon the initial state, say
\begin{eqnarray}
   \ketm{\phi}_{T} & = & \ketm{\phi}_{fin}
                   \;=\; \widehat{U}^{\:T}\: \ketm{\phi}_{int} \, .
\end{eqnarray}
In general, it is possible to start at position $\ketm{0}_{P}$ with any initial state
\begin{eqnarray}
   \ketm{\phi}_{int} & = & \left( a \ketm{0}_{C} \:+\:  b\ketm{1}_{C}\right) \otimes \ketm{0}_{P},
\end{eqnarray}
and with $a$ and $b$ just satisfying the normalization condition,
$\left\vert a\right\vert^{2} \:+\: \left\vert b\right\vert^{2}
\:=\:1$. For the sake of simplicity, we shall employ below the
following initial states to perform our QW
\begin{eqnarray*}
   \ketm{\phi}_{int} & = & \ketm{0}_{C} \otimes \ketm{0}_{P},                             \\[0.3cm]
   \ketm{\phi}_{int} & = & \ketm{1}_{C} \otimes \ketm{0}_{P}\, .
\end{eqnarray*}

The coin-shift operator can be applied to these initial states and gives rise, independent
of the particular values of $\theta $ and $T$ to
\begin{widetext}
\begin{eqnarray}
\left\vert 0\right\rangle _{C}\otimes \left\vert 0\right\rangle _{P}
&&\overset{1th}{\Longrightarrow }\sin \theta \left\vert 1\right\rangle
_{C}\otimes \left\vert -1\right\rangle _{P}+\cos \theta \left\vert
0\right\rangle _{C}\otimes \left\vert 1\right\rangle _{P} \nonumber \\
&&\overset{2th}{\Longrightarrow }\cos 2\theta \sin \theta \left\vert
1\right\rangle _{C}\otimes \left\vert -2\right\rangle _{P}+\sin 2\theta
\lbrack \sin \theta \left\vert 0\right\rangle _{C}+\cos \theta \left\vert
1\right\rangle _{C}]\otimes \left\vert 0\right\rangle _{P}+\cos 2\theta \cos
\theta \left\vert 0\right\rangle _{C}\otimes \left\vert 2\right\rangle _{P}...
\nonumber \\
&&\overset{Tth}{\Longrightarrow } \left\vert \phi \right\rangle _{fin}=\left( \prod\limits_{n=1}^{T}\cos n\theta
\right) \left\vert 0\right\rangle _{C}\otimes \left\vert T\right\rangle
_{P}+....+\left( -1\right) ^{T+1}\left( \prod\limits_{n=1}^{T}\frac{\cos n\theta }{%
    \cos \theta }\right) \sin \theta \left\vert 1\right\rangle _{C}\otimes
\left\vert -T\right\rangle _{P}, \label{A1}
\end{eqnarray}
\end{widetext}
\begin{widetext}
\begin{eqnarray}
\nonumber \\
\left\vert 1\right\rangle _{C}\otimes \left\vert 0\right\rangle _{P}
&&\overset{1th}{\Longrightarrow }\cos \theta \left\vert 1\right\rangle
_{C}\otimes \left\vert -1\right\rangle _{P}+\sin \theta \left\vert
0\right\rangle _{C}\otimes \left\vert 1\right\rangle _{P} \nonumber \\
&&\overset{2th}{\Longrightarrow }\cos 2\theta \cos \theta \left\vert
1\right\rangle _{C}\otimes \left\vert -2\right\rangle _{P}+\sin 2\theta
\lbrack -\cos \theta \left\vert 0\right\rangle _{C}+\sin \theta \left\vert
1\right\rangle _{C}]\otimes \left\vert 1\right\rangle _{P}+\cos 2\theta \sin
\theta \left\vert 0\right\rangle _{C}\otimes \left\vert 2\right\rangle _{P}...
\nonumber \\
&&\overset{Tth}{\Longrightarrow }\left\vert \phi \right\rangle _{fin}=\left( \prod\limits_{n=1}^{T}\frac{\cos
    n\theta }{\cos \theta }\right) \sin \theta \left\vert 0\right\rangle
_{C}\otimes \left\vert T\right\rangle _{P}+....+\left( -1\right) ^{T}\left(
\prod\limits_{n=1}^{T}\cos n\theta \right) \left\vert 1\right\rangle
_{C}\otimes \left\vert -T\right\rangle _{P}. \label{A2}
\end{eqnarray}
\end{widetext}
From expressions (\ref{A1}) and (\ref{A2}), we find that the
walker generally occupies $T+1$ positions after $T$ steps, i.e.
there is a non-zero probability at $T+1$ positions. A maximum of
$2^{T}$ terms may arise after $T$ steps for occupied positions for
the wave function of the walker. For an \textit{odd} number of
steps, only the odd positions will have non-zero probability
density, while the same is true for the even positions after an
\textit{even} number of steps. Moreover, the walker can generally
occupy only positions within $\left[-T,T\right]$. In the next
section, we shall make use of these properties to further classify
the behavior of walker, and hence its probability density
distribution in position space, if different rotation angles are considered.

\section{Simulation of the walk}     \label{Simulation}

\subsection{Classification}     \label{Classification}

The stepwise evolution (Eqs. (\ref{A1}) and (\ref{A2})) of the QW
shows that the state and probability distribution of the walker
sensitively depend on the particular choice of the rotation angle
$\theta$. Below, we shall therefore highlight the possible classes 
of behavior for the walker with different choices of $\theta$.
We compare their behaviors with those of a \textit{step-independent coin} (SIC) in order to better understand the effects of a \emph{step-dependent coin} (SDC) upon the
evolution of a walk. We focus on the wave function and the
probability density distribution in position space. All the simulations are done for
the initial state of $\ketm{0}_{C} \otimes \ketm{0}_{P}$.

\textbf{\textit{Localized walk:}} In this class, the probability 
density distribution is whether \textit{completely localized} at some position and/or 
it is periodically distributed only over two positions. In other words, at each step, the walker is relocated from one position into another without any distribution in its
wave function. This property of \textit{relocation} and absence of
the distribution in wave function remain unchanged for arbitrary
number of steps. In general, this class contains three subclasses; \textit{I) Free localized walk:} in which after $T$ steps of the walk, the probability density of the walker at
position $T$ is $1$ while at other positions, it is zero. An example of this case is $\theta=0$ (see Figs. \ref{0P-1} and \ref {GP0}). \textit{II) Bounded localized walk:} where the relocation of walker is limited to some specific range of positions. The walker starts at
position $0$ and after arbitrary steps, it returns to position $0$ again. In contrast to the previous case, the probability distribution of the walker \textit{oscillates} at different and within a specific
range of positions. The rotation angle of $\theta=\pi/2$ is an example of this walk (see Figs. \ref{P2-1} and \ref {GP2}). \textit{II) Bounded localized walk with periodic splitting:} here, at particular number of steps, the probability density distribution is completely localized at one position. While at other step numbers, the probability density of the walker is
distributed over two positions. The walk is limited to specific range of positions and after several steps, the same behavior and probability density distributions are repeated. Using coins with $\theta=\pi/4 \text{ and } \theta=\pi/6$ results into this behavior (see Figs. \ref {P4-1} and \ref {GP4}). 

\textbf{\textit{Classical like walk:}} The behavior of walker for this class results into \textit{Gaussian like} probability density distribution. The speed and the variance of walker is relatively smaller than usual quantum random walk and similar to CW. There are two distinctive subclasses for this class; \textit{II) Compact classical walk:} in which a Gaussian like probability density distribution is found for the walker. But, unlike the CW, probability density is distributed over just a very few number of positions and could not exceed this range of positions for arbitrary number of steps (it is bounded). This results into a steeper Gaussian like behavior, hence \textit{compact classical like} walk. Coin with $\theta=\pi/12$ leads to such behavior (see Figs. \ref{P12-1} and \ref {GP12}). \textit{II) Classical like walk:} where in contrast to previous case, the Gaussian like probability density distribution of the walker is not bounded to specific range of positions and it is an increasing function of steps. The rotation angle of $\theta=3.59\pi /5$ is an example of this case (see Fig. \ref{15P12-1} and Fig. \ref {G15P12}). The justification for this classification and its subclasses are provided in \ref{Classic}. 

\textbf{\textit{Semi-classical/quantum like walk:}} In this class, the walker enjoys some characteristics of the both quantum and classical walks. There is a significant peak in the probability density distribution at center of position (or very close to it). But, other noticeable peaks are also observed at left and/or right hand side of it. This results into a multimodal distribution for the probability density. In general, it is hard to classify this distribution into more subclasses. This is because each example of this class has specific properties (similar to classical or quantum walk) which are different from the other examples. But in general, they have similarities to the cases where decoherence is present in the QW. For example, in case of $\theta=2\pi/5$, observed probability density distribution has the highest peak at the center of position with two other smaller peaks at left and right hand sides of the distribution (see Figs. \ref{2P5-1} and \ref {G2P5}). This is similar to the case where system has decoherent coin (for more details, please see \ref{Semi}). Another example for this class is walk with a coin characterized by $\theta=\pi/5$. A multimodal distribution is observed for the probability density (see Figs. \ref{P5-1} and \ref {GP5}). 

\textbf{\textit{quantum like walk:}} For this class, the walker could obtain a ballistic asymmetric like distribution in its probability density. The highest peak in probability density is at right/left hand side of position's center and its place is an increasing function of the steps. Fast spread of the walker's wave function and large variance are other characteristics of this class. These are the properties of usual QW. Coin with rotation angle of $\theta=\pi/3$ results into such walk (see Figs. \ref{P3-1} and \ref {GP3}).

\begin{samepage}
	
\begin{figure*}[!htbp]
	\renewcommand*\thesubfigure{\arabic{subfigure}}
	\centering
	{\begin{tabular}[b]{ccc}
		\subfloat[$\theta=0$]{\label{0P-1}\includegraphics[width=0.3\textwidth]{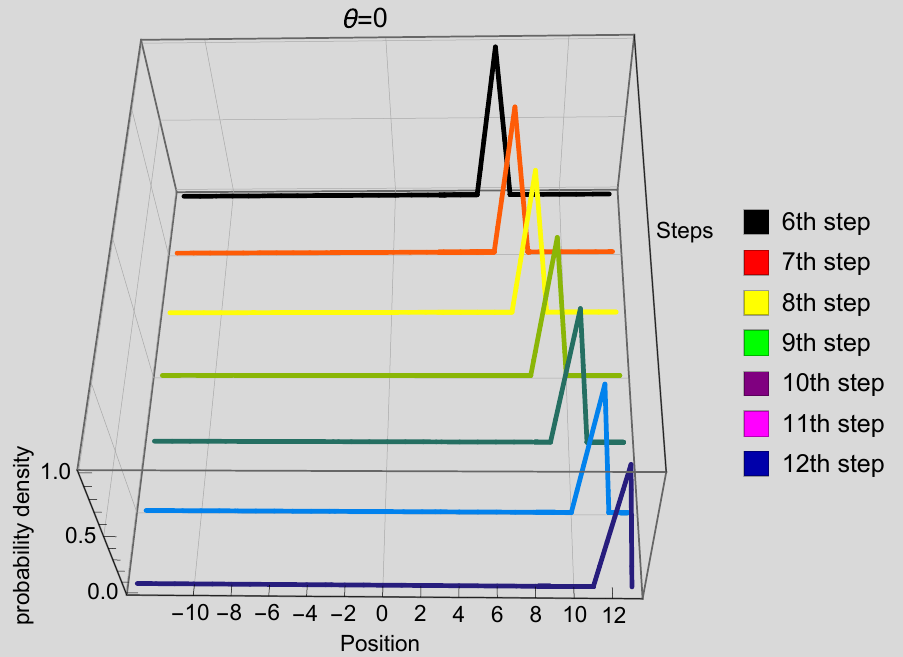}}
		\quad 
		\subfloat[$\theta=\pi/2$]{\label{P2-1}\includegraphics[width=0.3\textwidth]{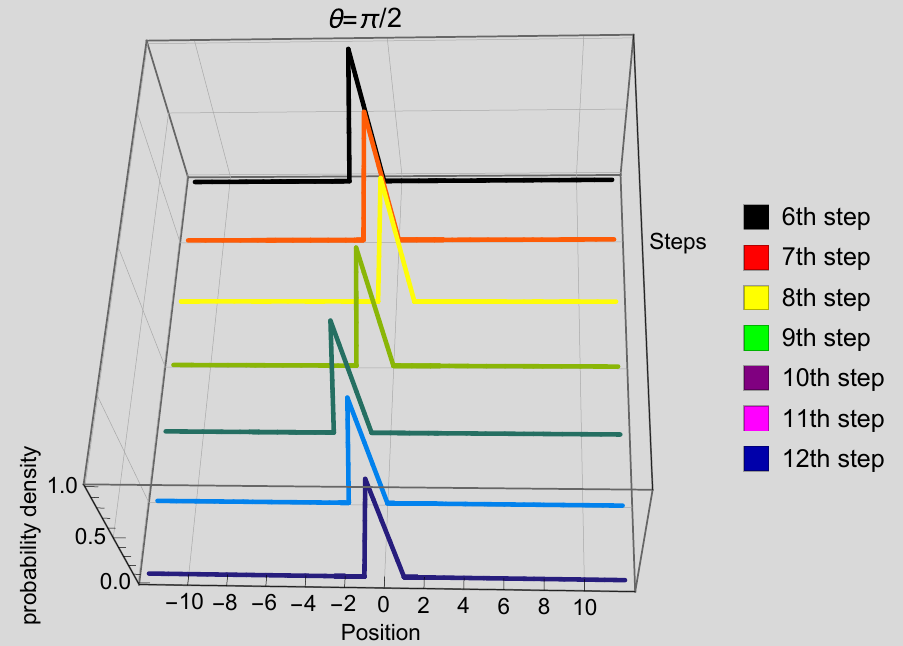}} \quad 
		\subfloat[$\theta=\pi/4$]{\label{P4-1}\includegraphics[width=0.3\textwidth]{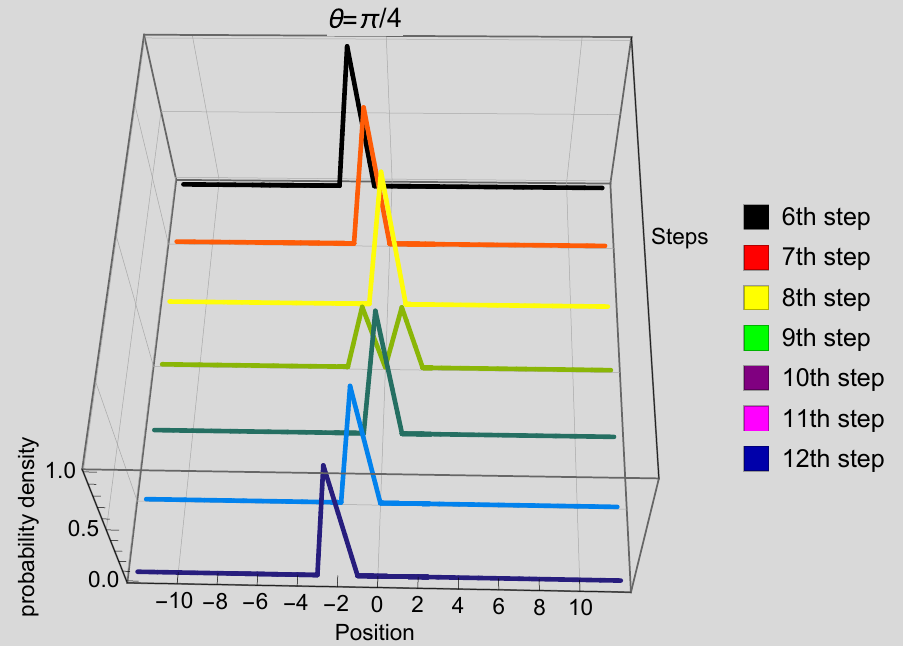}} 	
		\\ [0.05cm]  	
		\subfloat[$\theta=\pi/12$]{\label{P12-1}\includegraphics[width=0.3\textwidth]{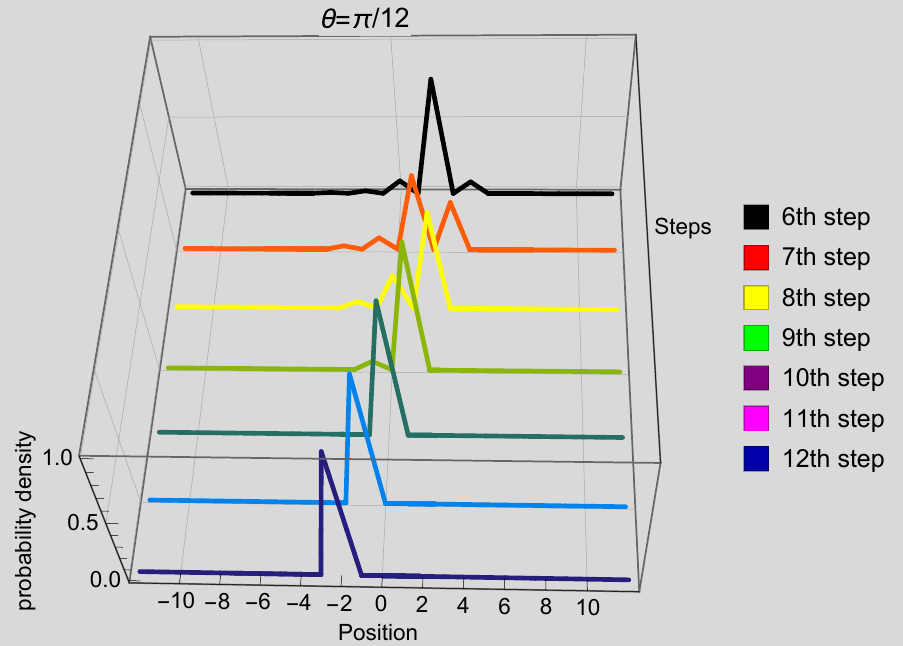}} \quad 
		\subfloat[$\theta=3.59\pi/5$]{\label{15P12-1}\includegraphics[width=0.3\textwidth]{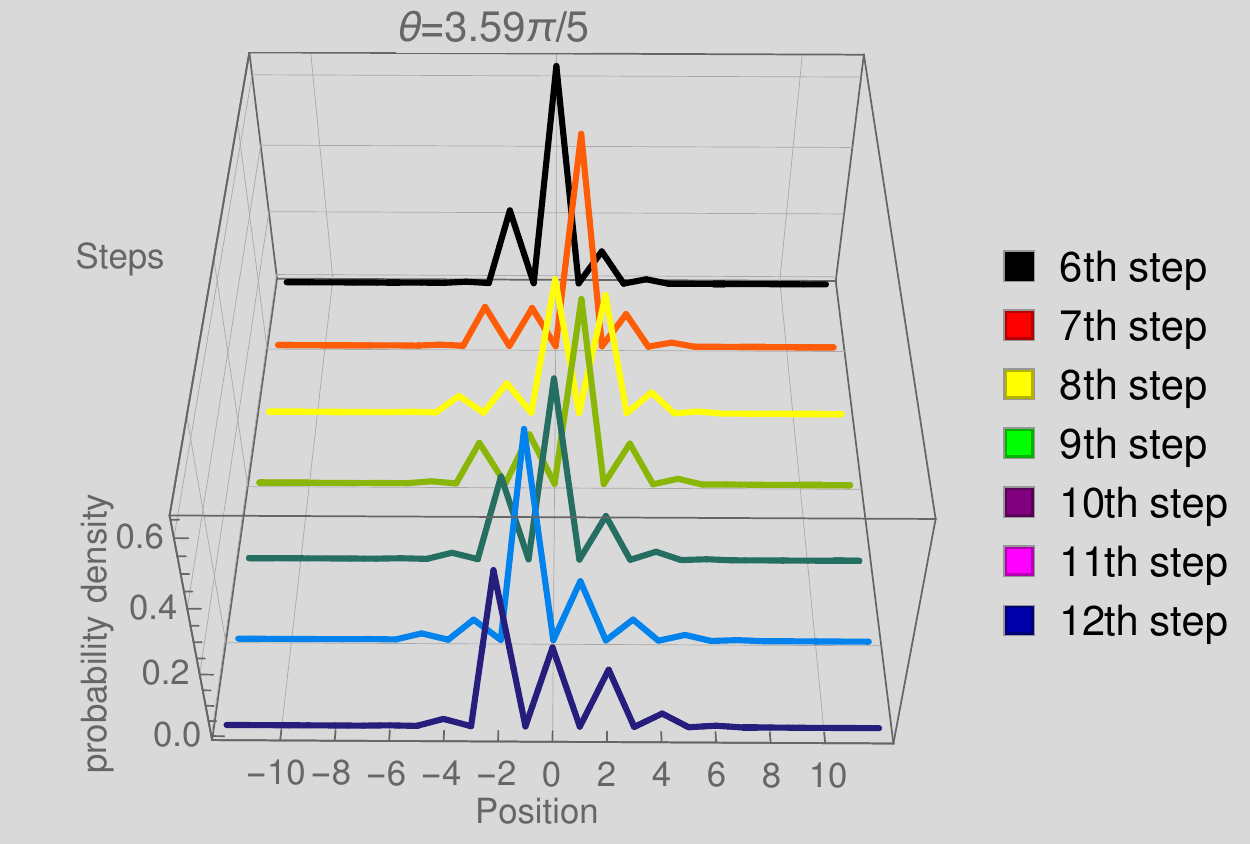}}  
		\quad 
		\subfloat[$\theta=2\pi/5$]{\label{2P5-1}\includegraphics[width=0.3\textwidth]{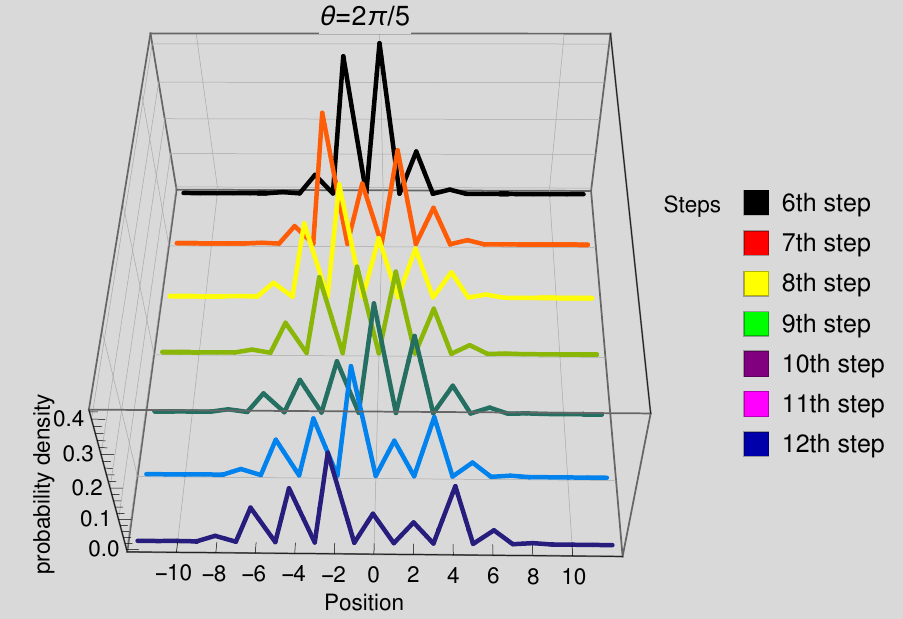}} 
		\\ [0.05cm]			
		\subfloat[$\theta=\pi/5$]{\label{P5-1}\includegraphics[width=0.3\textwidth]{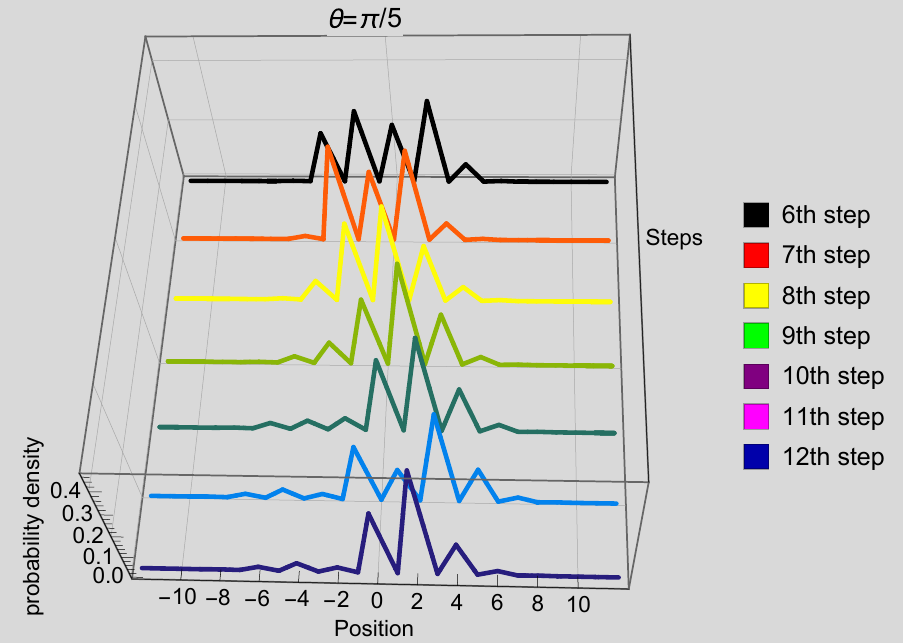}} 	   \quad 
		\subfloat[$\theta=\pi/3$]{  \label{P3-1}\includegraphics[width=0.3\textwidth]{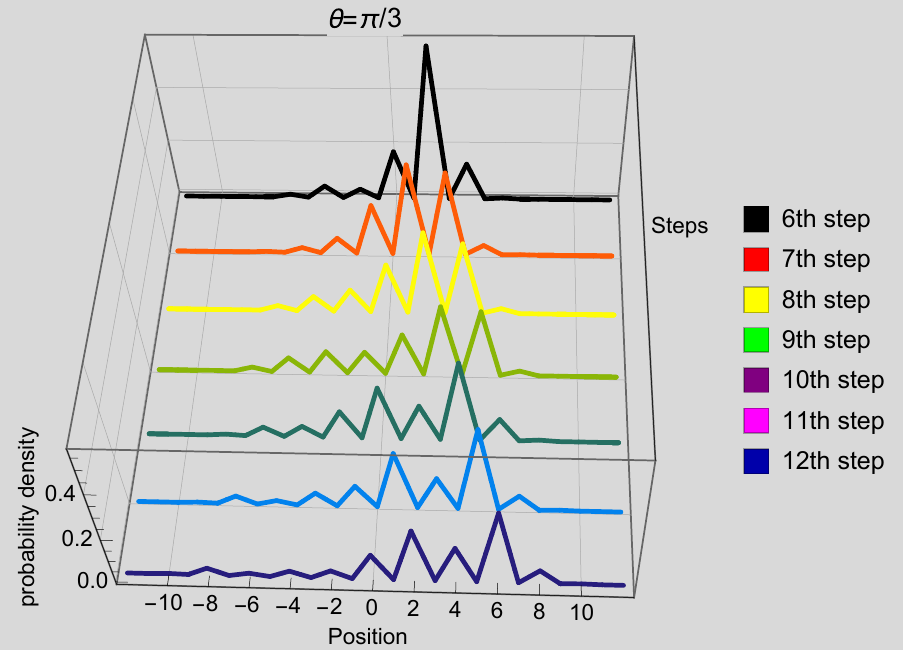}}
	\end{tabular}} 	
	\caption{Probability density versus position for different rotation angles. Results are shown for seven subsequent steps $T = 6..12$.}
	\label{Fig1}
\end{figure*}
		
\begin{figure*}[!htbp]
	\centering
		\includegraphics[width=0.32\linewidth]{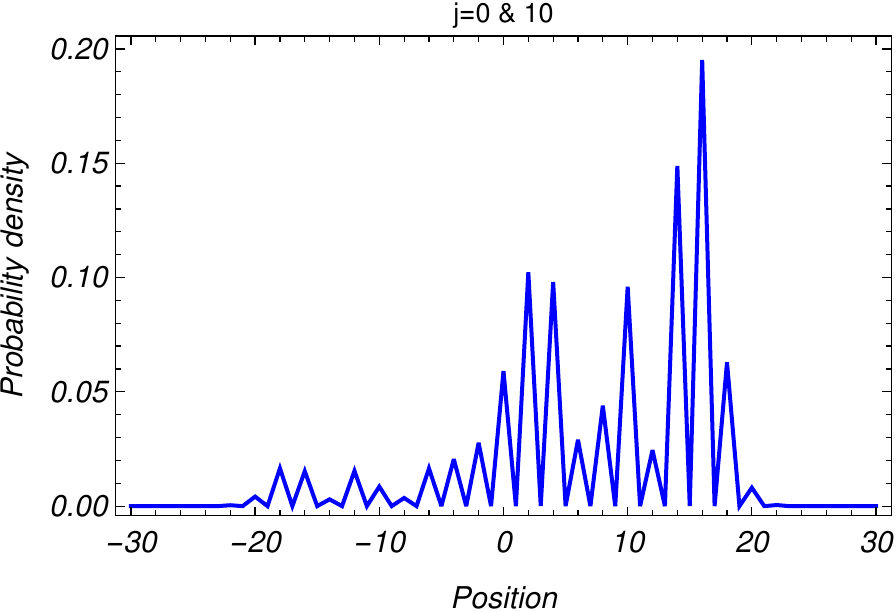}  \label{0} \hfil
		\includegraphics[width=0.32\linewidth]{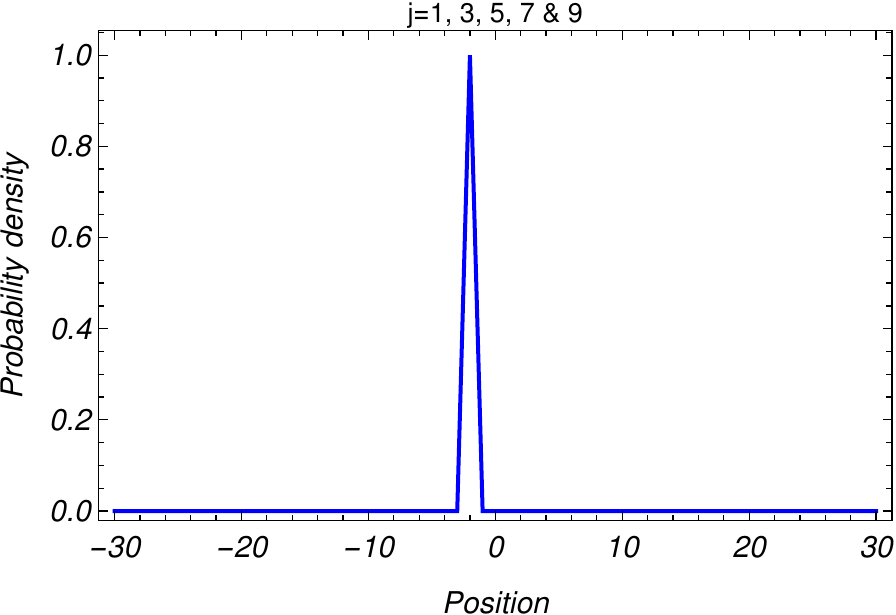}  \label{1}
		\par \medskip
		\includegraphics[width=0.32\linewidth]{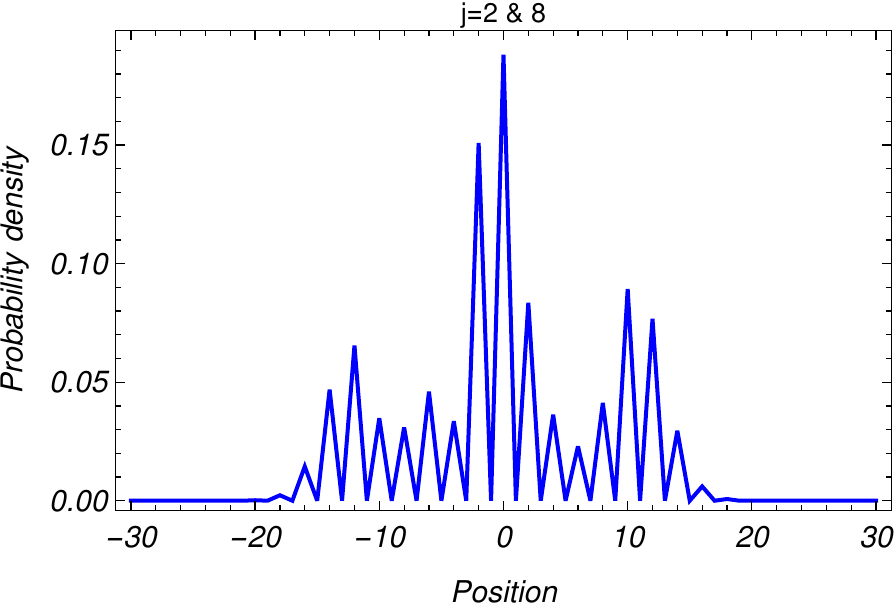}  \label{2} \hfil
		\includegraphics[width=0.32\linewidth]{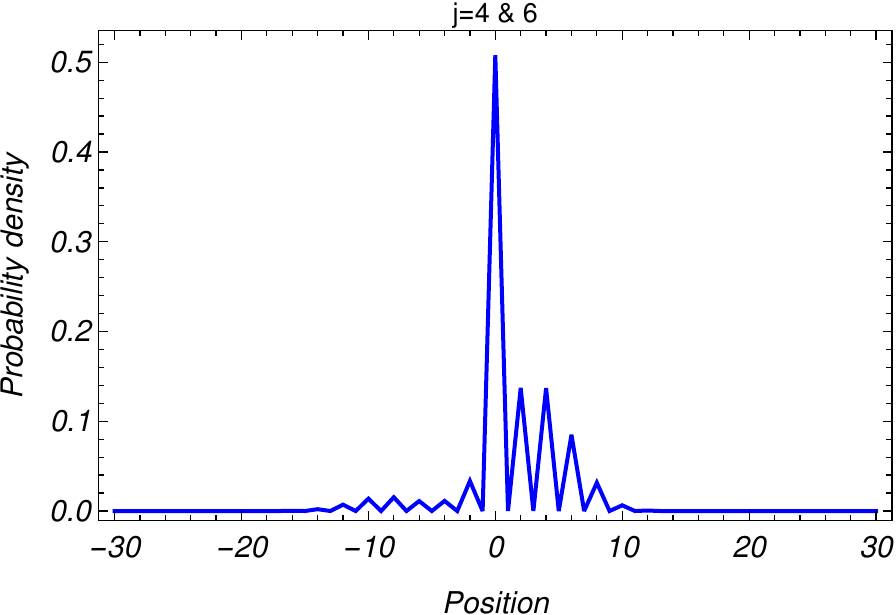}  \label{3}
		\caption{Probability density versus position for $\theta^{'}=
				\theta+0.j \theta$ where $\theta=\pi/3$.}
		\label{Fig2}
\end{figure*}		
\end{samepage}

The different behaviors of a QW with a step-dependent coin (\ref{coin}) are summarized again in Table~I.
\begin{table}[ht]
    \centering
    \caption{Classification of observed probability density distribution for walk with step dependent coin.}
\label{table}
\begin{tabular}{| l | r |}
        \hline
        \textbf{Classes}              & \textbf{$\theta$} \\
        \hline
        Localized: free                             & $0$ \\
        \hline
        Localized: bounded                          & $\pi /2$   \\
        \hline
        Localized: bounded with periodic splitting   & $\pi /4$, $\pi /6$  \\
        \hline
        Compact classical like            & $\pi /12$  \\
        \hline
        Classical like            & $3.59\pi /5$  \\
        \hline
        Semi-classical/quantum like       & $\pi /5$, $2\pi /5$   \\
        \hline
        Quantum like                      & $\pi /3$   \\
        \hline
\end{tabular}
\end{table}

\begin{figure}[!htb]
        \centering
        \includegraphics[width=0.7\linewidth]{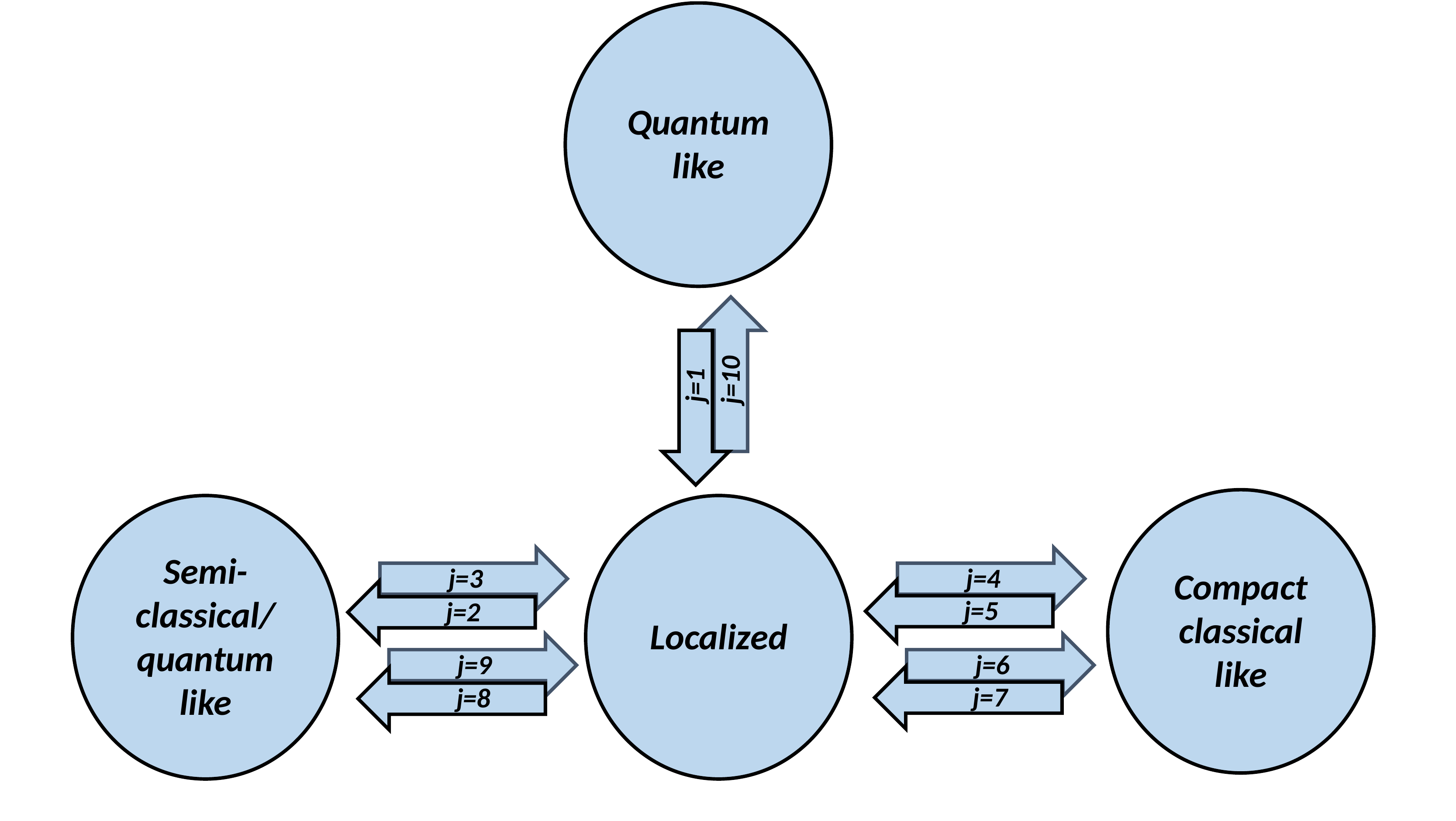}
        \caption{Modification in the walker’s behavior for $\theta^{'}=\theta+0.j \theta$ whit $\theta=\pi/3$ and $j=1,2,...,10$.}
        \label{Fig111}
    \end{figure}

Next, we study walks with a rotation angle $\theta^{'}= \theta+0.j
\theta$ in which $j$ is an integer running between $0$ and $10$.
We choose $\theta=\pi/3$ (see Fig. \ref{Fig2}). The goal
is to examine how the QW behaves for different rotation angles $\theta$ for specific values between
$\theta=\pi/3$ to $\theta=2\pi/3$. For
$\theta=\pi/3$, the walker exhibits a \emph{quantum like}
probability density. For odd integers of $j$, the wave function is
completely \textit{localized} at position $-1$ while an identical
\emph{semi-Gaussian like} distribution with a peak at zero is found for
$j=2$ and $j=8$. One can categorize this distribution under
\textit{semi-classical/quantum like} behavior. For $j=4$ and $6$,
the distribution of probability density is also identical and it
shows \textit{classical like behavior} (similar to the
$\theta=3.59\pi /5$ case). Finally, probability density
distribution of the case $\theta=2\pi/3$ is identical to
the one observed for $\theta=\pi/3$. To summarize, we can
see that from $\theta=\pi/3$ to $\theta=2\pi/3$, probability density 
distribution will be modified as shown in Fig. \ref{Fig111}.

It is worthwhile to mention that the probability density distributions of $\theta$ and $\pi-\theta$ are always identical.
\subsection{Step dependent vs. step independent}    \label{Step dependent}

\begin{figure*}[!htbp]

	\centering
	\renewcommand{\thesubfigure}{\alph{subfigure}}
	\subfloat[Probability density versus position. \label{Probab}]
	{\begin{tabular}[b]{cccc}%
	\setcounter{subfigure}{0}		
	\renewcommand*\thesubfigure{\arabic{subfigure}}			
	\subfloat[$\theta=0$]{\includegraphics[width=0.26\textwidth]{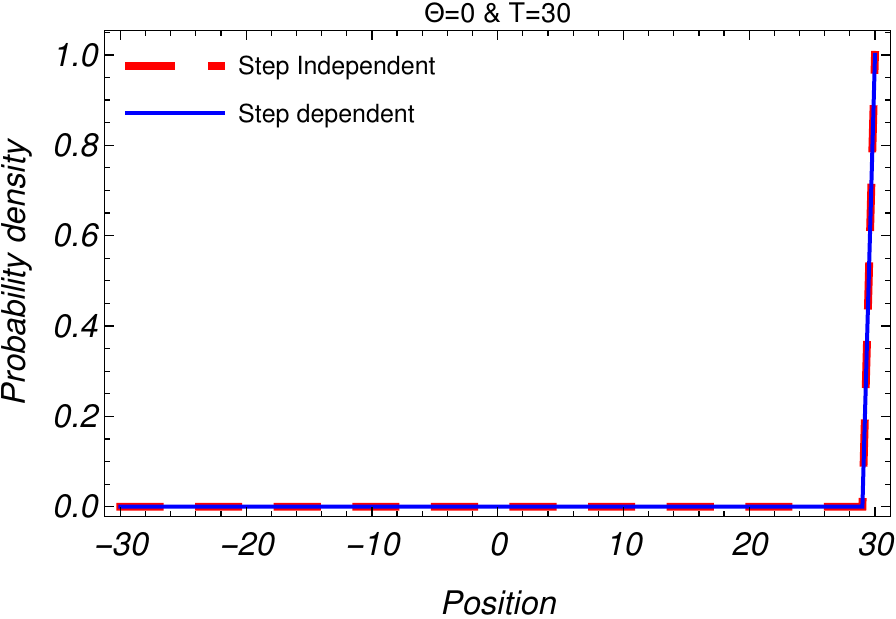}} \label{0P-Compare}	
	\subfloat[$\theta=\pi/2$]{\includegraphics[width=0.26\textwidth]{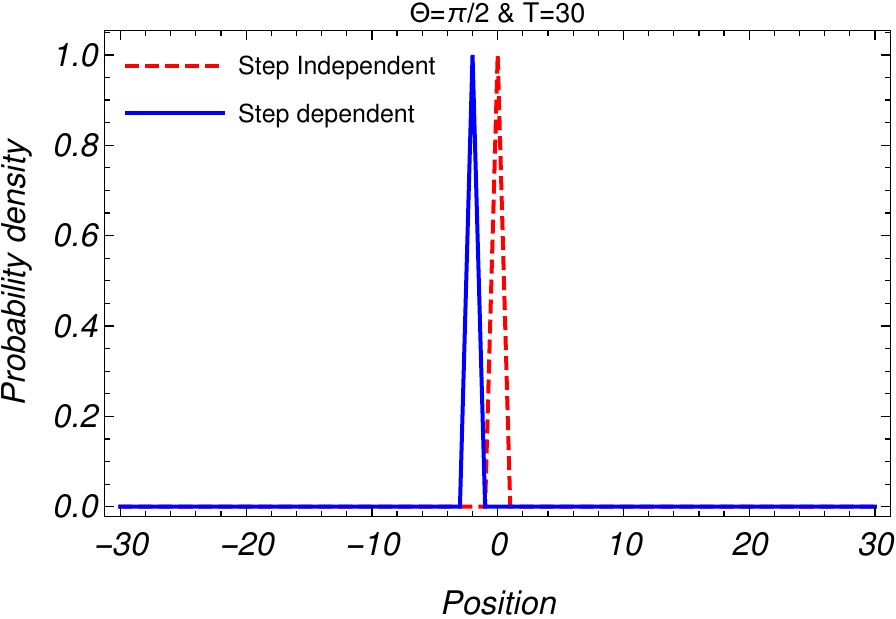}} \label{P2-Compare}
    \subfloat[$\theta=\pi/4$]{\includegraphics[width=0.26\textwidth]{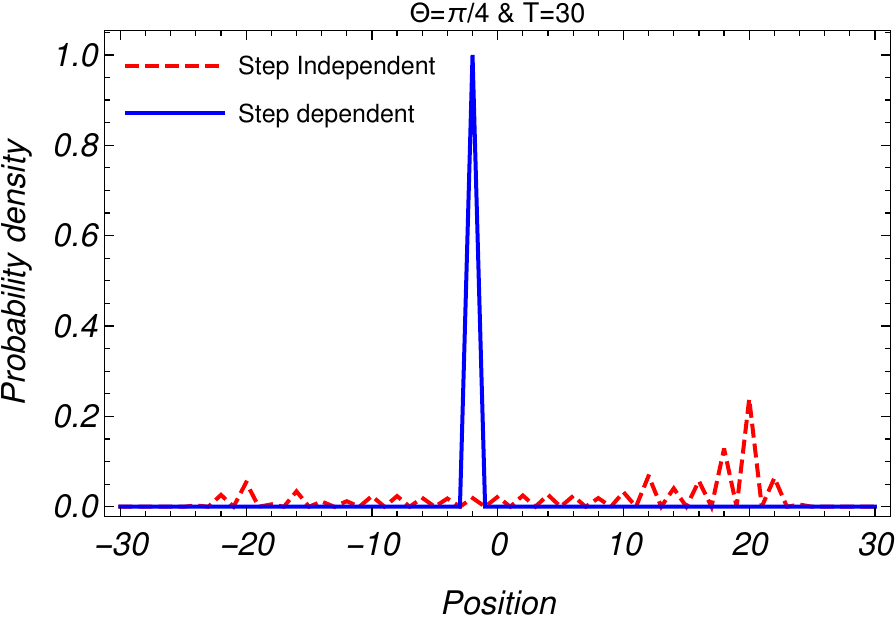}} \label{P4-Compare}
	\subfloat[$\theta=\pi/12$]{\includegraphics[width=0.26\textwidth]{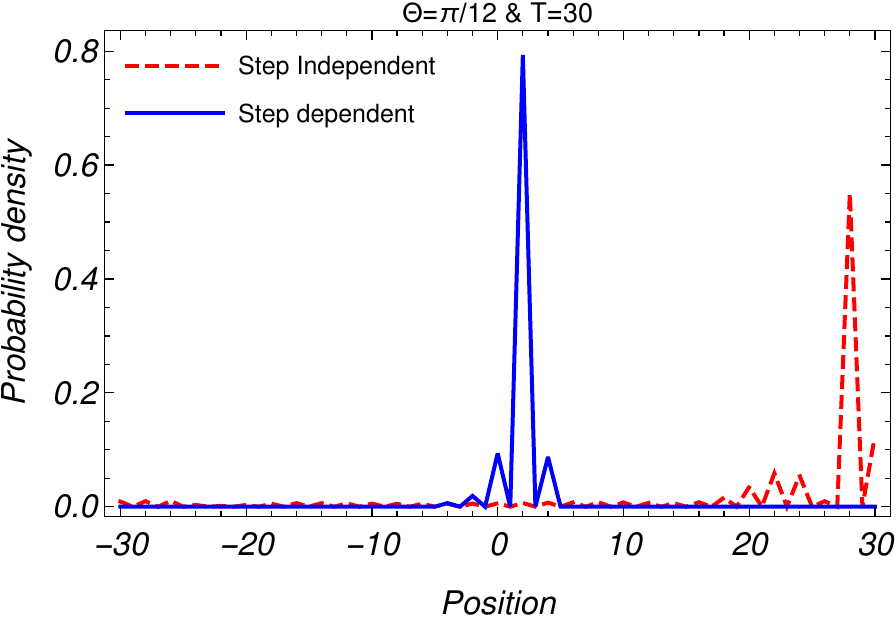}}  \label{P12-Compare}
	\\[0.5cm]
	\renewcommand*\thesubfigure{\arabic{subfigure}}
	\subfloat[$\theta=3.59\pi /5$]{\includegraphics[width=0.26\textwidth]{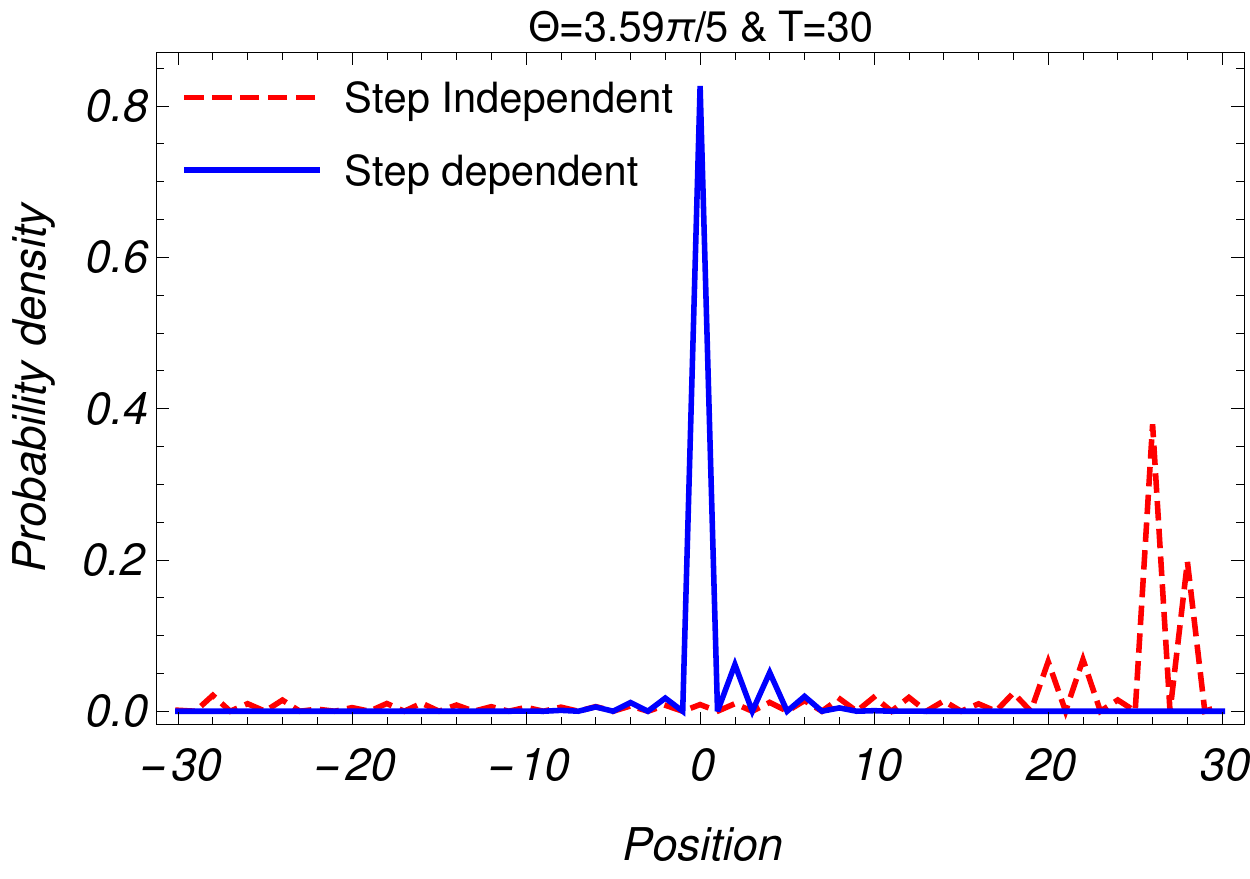}} \label{15P12-Compare} 	
	\subfloat[$\theta=2\pi/5$]{\includegraphics[width=0.26\textwidth]{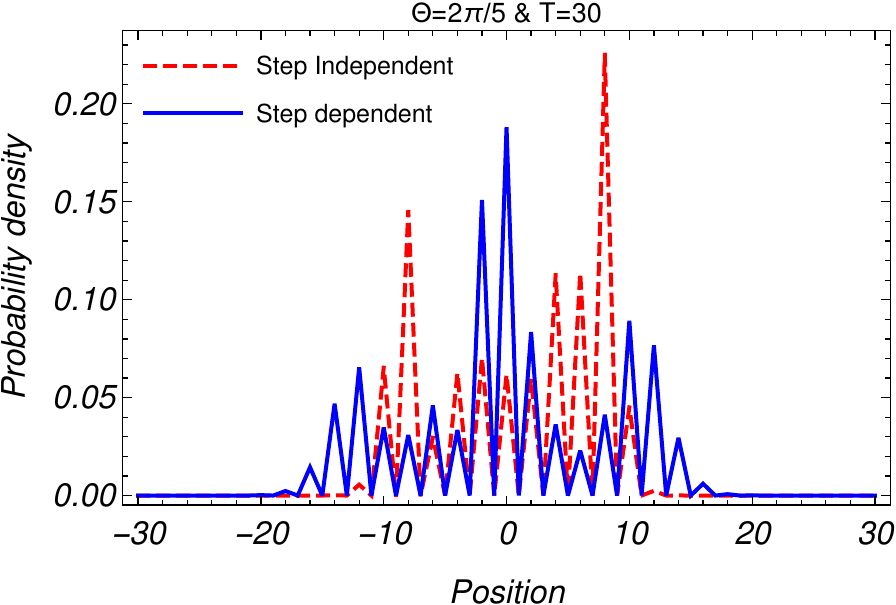}} \label{2P5-Compare} 
	\subfloat[$\theta=\pi/5$]{\includegraphics[width=0.26\textwidth]{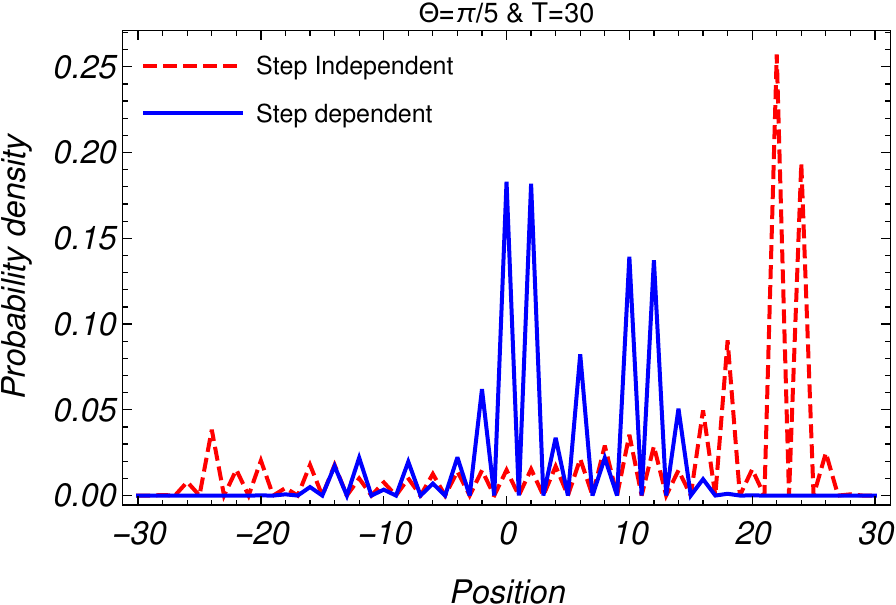}} \label{P5-Compare} 
	\subfloat[$\theta=\pi/3$]{\includegraphics[width=0.26\textwidth]{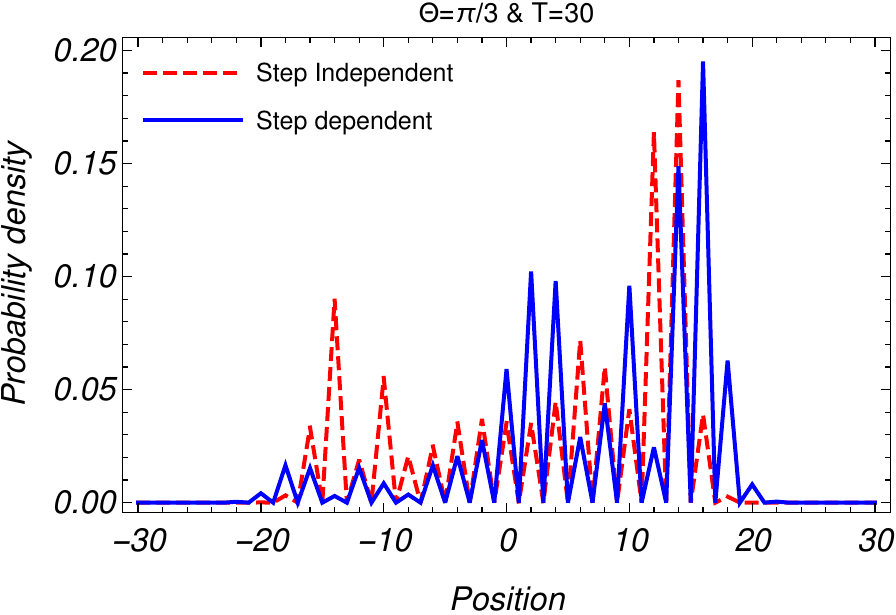}}	\label{P3-Compare}  
	\end{tabular}}	 
	\\ [0.05cm]
	
	\renewcommand{\thesubfigure}{\alph{subfigure}}
	\subfloat[Number of positions with non-zero probability density versus step; probability densities smaller than $10^{-4}$ are neglected.  \label{Variance}] 
	{\begin{tabular}[b]{ccc}%
	\setcounter{subfigure}{8}		
	\renewcommand*\thesubfigure{\arabic{subfigure}}			
	\subfloat[$\theta=\pi/3$ (continous line) and $\theta=\pi/4$ (dashed line)]{\includegraphics[width=0.3\textwidth]{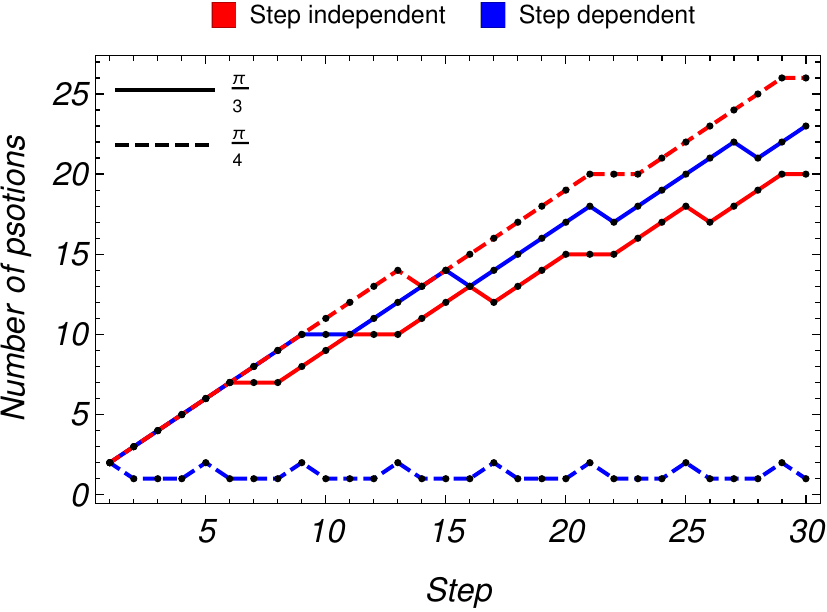}} \label{N1}	
	\subfloat[$\theta=\pi/5$ (continous line) and $\theta=\pi/12$ (dashed line)]{\includegraphics[width=0.3\textwidth]{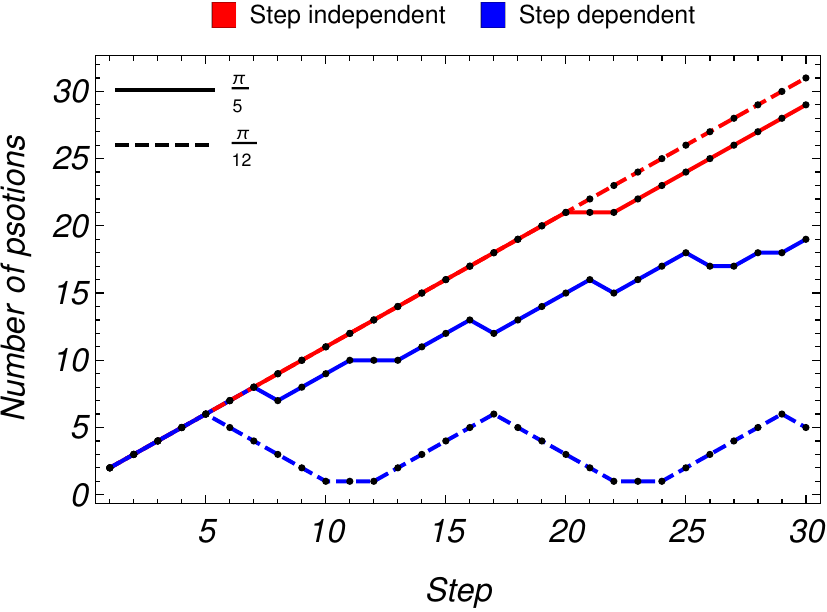}} \label{N2} 
	\subfloat[$\theta=2\pi/5$ (continous line) and $\theta=3.59\pi /5$ (dashed line)]{\includegraphics[width=0.3\textwidth]{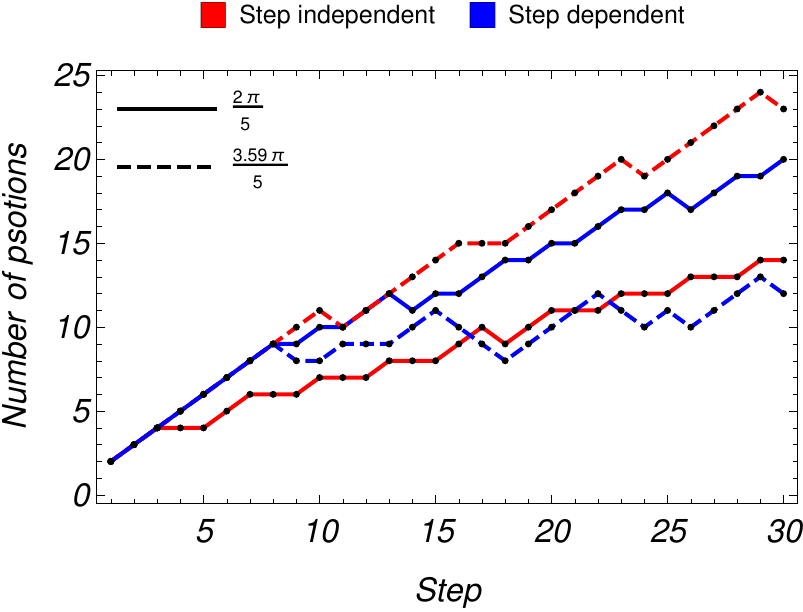}} \label{N3} 
	\end{tabular}} \\ [0.05cm]
	
	\renewcommand{\thesubfigure}{\alph{subfigure}}	
    \subfloat[Entropy versus steps. \label{Entropy}] 
	{\begin{tabular}[b]{ccc}%
	\setcounter{subfigure}{11}
    \renewcommand*\thesubfigure{\arabic{subfigure}}			
    \subfloat[$\theta=\pi/4$]{\includegraphics[width=0.3\textwidth]{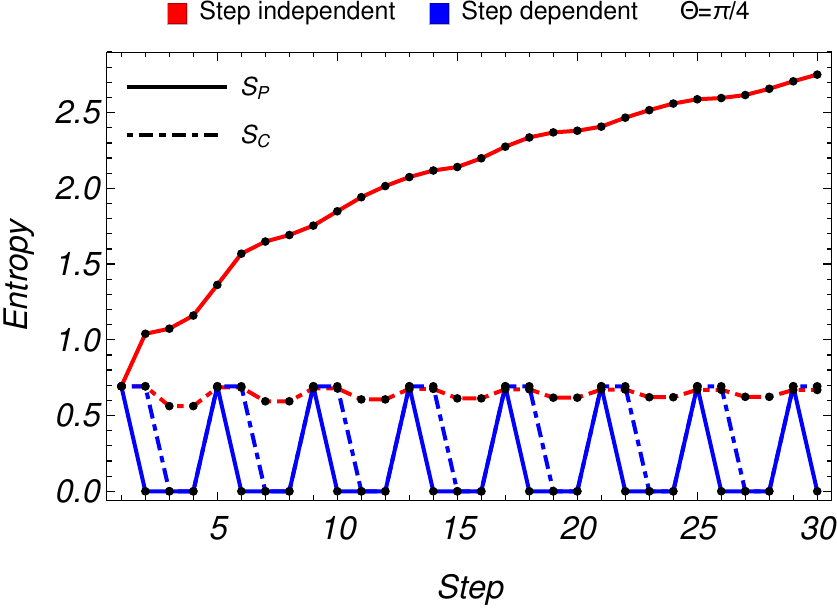}} \label{E2}  	
	\subfloat[$\theta=\pi/12$]{\includegraphics[width=0.3\textwidth]{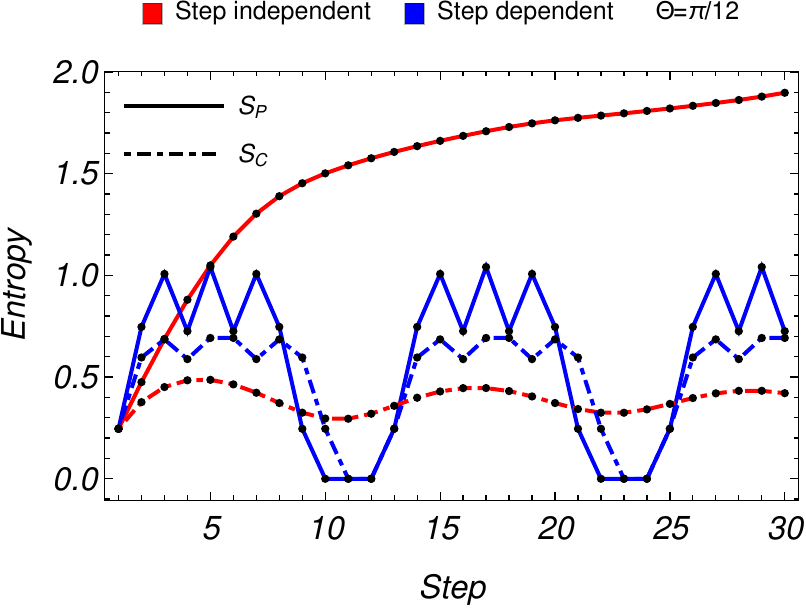}} \label{E5}
	\subfloat[$\theta=3.59\pi /5$]{\includegraphics[width=0.3\textwidth]{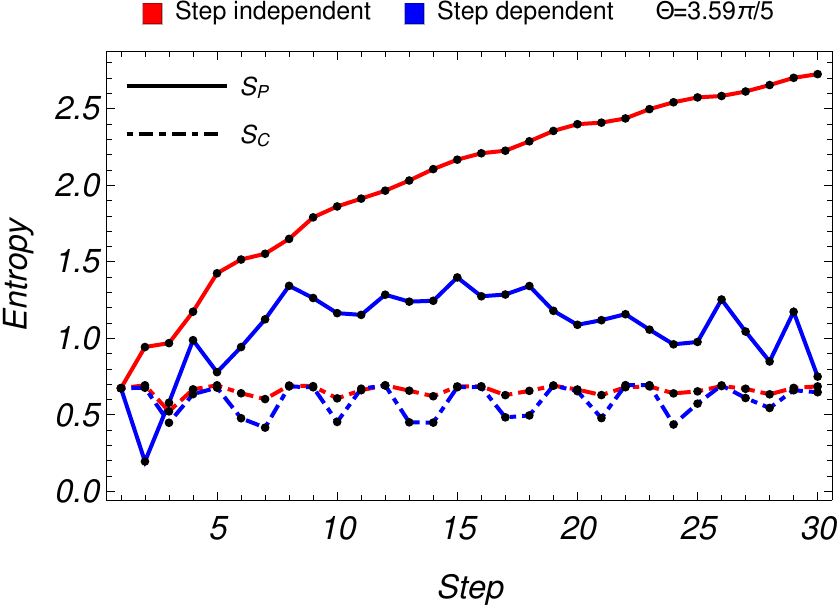}} \label{E6} \\[0.01cm]
    \renewcommand*\thesubfigure{\arabic{subfigure}}	
	\subfloat[$\theta=2\pi/5$]{\includegraphics[width=0.3\textwidth]{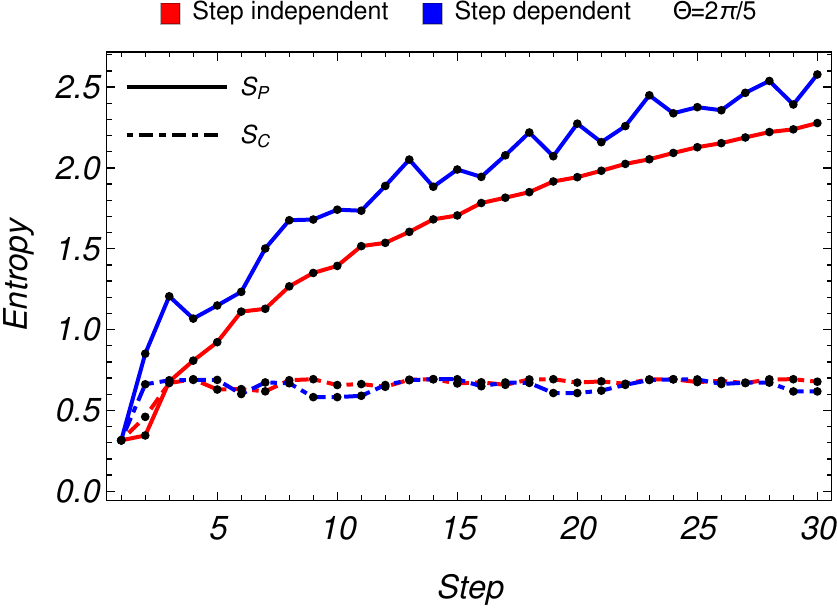}}{\label{E4}} 	
	\subfloat[$\theta=\pi/5$]{\includegraphics[width=0.3\textwidth]{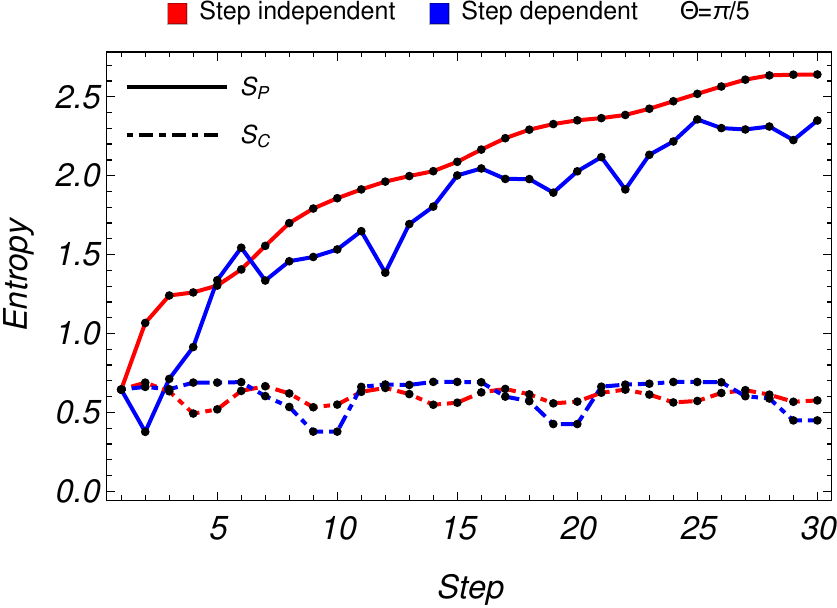}}  \label{E3}
	\subfloat[$\theta=\pi/3$]{\includegraphics[width=0.3\textwidth]{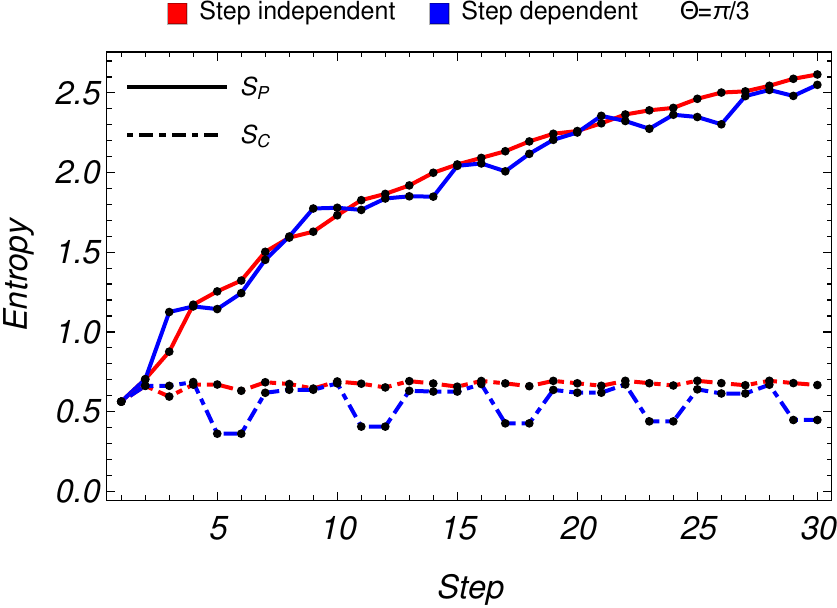}}  \label{E1}
	\end{tabular}}   \\
	\caption{Step-dependent coin (blue colored diagrams) versus step-independent coin (red colored diagrams)} \label{Fig3}
\end{figure*}

When compared with a SIC walk, we find the following behaviors for
QW with a SDC; for $\theta=0$, both SDC and SIC
walks have identical distribution for probability density (see Fig.
\ref{Fig3} --1). As for the rotation angle of $\theta=\pi/2$,
completely localized probability density distribution is observed
for both SDC and SIC but at different positions (see Fig.
\ref{Fig3} --2). As for the $\theta=\pi/3$, we notice two
significant differences between SDC and SIC cases: I) The variance
of SDC walk is larger than SIC one. II) The SIC has an asymmetric
probability density distribution with sharp peaks at right and left hand sides of the
distribution (not at the center). Whereas for SDC, there is a peak
at the center of position and probability density of the highest
peak is larger comparing to SIC walk (see Fig. \ref{Fig3} --8).

A significant difference between SDC and SIC walks could be seen
for $\theta=\pi/4$. For the SIC, this coin is known as
Hadamard. The wave function with Hadamard coin is asymmetrically
distributed over $[-T/\sqrt{2},T/\sqrt{2}]$
positions \cite{Kempe}. Whereas, for the SDC walk, the complete
localization with periodic $50-50$ distribution over two positions
takes place for the wave function. In addition, the distribution
of wave function of the SIC case is an increasing function of the
steps while for the SDC, it is limited to specific number of the
positions (see Fig. \ref{Fig3} --3). As for $\theta=\pi/5$,
the SIC provides wave function with larger number of the occupancy
over positions comparing to SDC. In addition, the highest
probabilities are located at left and right hand sides of distribution in SIC
walk, whereas, the SDC has a multimodal distribution (see
Fig. \ref{Fig3} --7). In case of $\theta=2\pi/5$, larger variance in probability density distribution is observed for SDC comparing to SIC walk. For the $\theta=\pi/12$, in
the SDC walk, variance of wave function is limited to specific
region while for the SIC, it is an increasing function of the
steps. On the other hand, the walk with SDC presents a compact Gaussian
like distribution in probability density, while the SIC enjoys an
asymmetric behavior (see Fig. \ref{Fig3} --4). The same properties could be stated for $\theta=3.59\pi /5$ with small modifications (Gaussian like distribution and unbounded variance) (see Fig. \ref{Fig3} --5).

Next, we study the number of positions with non-zero probability
density as a function of steps for SDC and SIC walks (see Fig.
Fig. \ref{Fig3} --k). Evidently, one can recognize the periodic behavior
and complete localization that were reported before for cases
$\theta=\pi/4 \text{ and } \pi/12$. For
characteristic rotation angles of $\theta=\pi/5 \text{ and } 3.59\pi /5$,
comparing to SIC walk, SDC walk has smaller number of positions
with non-zero probability density. This confirms the classical
like property that was mentioned for these cases. In contrast, SDC
walks with $\theta=\pi/3 \text{ and } 2\pi/5$ have higher number of positions
with non-zero probability density comparing to SIC walk. This
shows that SDC walk is faster than SIC walk for these rotation angles.

\subsection{Discussion on physical interpretations}   \label{Discussion}

A localized probability density distribution may signal the
presence of Anderson localization. About 60 years ago, Anderson
predicted that destructive interferences among scattering paths of
a quantum particle propagating in a disordered media may occur and
results into a localization of the wave function \cite{Anderson}.
Such localizations have been extensively investigated in different
physical contexts such as photons and photonic lattices
\cite{Lahini}, Bose-Einstein condensate \cite{Roati}, microwaves
in strongly scattering samples \cite{Gross} and inverted opals
\cite{Conti}.

In the context of QW, an Anderson localization could be realized
as a case where the mean-squared distance of the walker from the
origin stays bounded even in the infinite-time limit. The
theoretical existence of such localization for the QW was pointed
out in Refs. \cite{Joye,Nicola}. In Ref. \cite{Rakovszky}, the Anderson
localization was observed in one-dimensional split-step quantum walk.  
This localization was due to introduction of static, symmetry-preserving disorder in the
parameters of walk. Experimentally, an Anderson localization
was observed for photonic implementation of a one
dimensional quantum walk in the presence of decoherence \cite%
{Broome,Schreiber2} and pairs of polarization-entangled photons in
a discrete quantum walk affected by position-dependent disorder
\cite{Crespi}. In all of the mentioned papers, the existence of
Anderson localization was shown to be rooted in the presence of
disorder in the QW. From our discussion above,
we see that a walk with a SDC can result into an Anderson
localization without any disorder
must occur. In other words, it was shown that a unitary evolution
of a walk without any special modification can result into probability density distributions
similar to the ones in the presence of disorder. This
shows that QW with a SDC can be utilized to realize the processes
such as Anderson localization in a more simplified framework. In
many experimental setups, moreover, it is simpler to use a SDC
walk to achieve Anderson localization and exploiting its
properties, instead of using disorder-included systems (see for
example Ref. \cite{Ying}).

Previously, it was shown that transition from QW into
classical one could be achieved by introducing the decoherence into system
through; I) Subjecting the decoherence into coin of system \cite%
{Brun,Annabestani}. II) Introduction of random phase shifts to the
coin particle \cite{kosik} III) Measurements performed with timing
provided by the Levi waiting time distribution \cite{Romanelli}.
All of these processes require introduction of decoherence into the
system. The SDC presented in this paper provides the possibility
of achieving classical like behavior without introduction of the
decoherence into system. This indicates that while we are
employing the unitary operators, it is possible to have classical
probability density distribution for the walker. This enables us to avoid
certain complex calculations (for more details please see Refs.
\cite{Kendon,Venegas-Andraca} and references within). This results
into a more simplified framework to study the quantum random walks
and its applications for classical like behavior \cite
{Schreiber,Schreiber2,Broome,Perets}.

Existence of different probability density distributions (ranging
from quantum like to complete localization) for SDC provides us
with a multi-purpose QW. Due to these different
classes of the probability density distribution, one can develop a
more adaptable quantum walk-based search algorithm \cite{Santha}. The main issue that must
be taken into account is the capability of coherent QW. An
isolated QW system (absence of decoherence), with certain
considerations (here, it is step dependency of the coin) has the
capability of providing large number of the possibilities for the
walker. This indicates that QW has the possibility of addressing
complicated simulations in the present literature in a more
simplified framework.

In Ref. \cite {Venegas-Andraca}, three key points are introduced
for building an algorithm based on quantum walks: I) The unitary
operators II) The measurement operators III) Decoherence effects
in order to control the quantum walk algorithmic effects by
manipulating probability distributions or mimicking natural
phenomena. Specially, in Ref. \cite{Mohseni}, the importance of
decoherence for manipulating probability density distribution of
the walker for mimicking natural phenomena was highlighted. As we
can see, SDC random walk has the possibility of addressing
decoherence-included QW. This indicates that for development of
algorithm, the third condition could be relaxed and only the two
first conditions are left. The first condition is also
automatically satisfied due to unitary nature of the coin-shift
operator of SDC walk. Therefore, only the second condition remains
to be addressed.

In quantum simulation, the main goal is providing quantum systems
which could be programmed to model the behavior of other quantum
systems \cite{Aspuru-Guzik,Mohseni}. Due to the diverse behaviors
and possibilities for SDC walk, it is clear that the SDC walk
could catch the idea of "programmable quantum system" in a better
suit. The general structure of the SDC provides a simplified
framework for manipulating the quantum system. The fact is that
SDC has a comfortable adaptability for providing different
probability density distributions which makes it ideal in terms of
programmability for quantum computation/simulation purposes.
Finally, the comparison for quantum like behavior case of SDC
versus SIC showed that the variance in SDC is bigger. This
indicates a faster spread of wave function of the walker for SDC.
This difference in the variance could be employed to conclude that
algorithms based on SDC could be faster comparing to their SIC
counterpart \cite{Kempe,Ambainis}. Therefore, quantum like
behaviors observed for SDC walks are showing better efficiency
regarding the hitting and mixing times comparing to SIC walks
\cite{Magniez}.

QWs can be used as universal generators of probability density distributions \cite{Montero3}. In Ref. \cite{Goyal}, the unitary equivalence of classes of QWs with different $SU(2)$ coin operators was shown and it was pointed out that the nonequivalent QWs can be distinguished by a single parameter. In fact, it was pointed out that walk with a dynamical coin is equivalent to electric QWs. Due to these properties and the fact that QW being a universal computational primitive, there has been a growing interest in using QWs as a tool for state engineering, initial state preparation and state transferring. For example, it was pointed out that using the localization like effects in QWs, one can engineer a state of walker and obtain specific desired details for its state \cite{Franco,Majury}. In Ref. \cite{Innocenti}, a beautiful theoretical framework was proposed for quantum state engineering of arbitrary superpositions of the walker’s sites. Experimentally, Nitsche et al used QWs with dynamical control to prepare non-localized input states and implement a state transfer scheme for an
arbitrary input state \cite{Nitsche}. In this direction, we can see that walks with a step-dependent coin can also be used as a stand alone framework for generating desired probability density distributions, preparing initial state and/or engineer them. The classification that we provided shows that each class actually is made out of different subclasses. For example, for localized walk, we showed that there are at least three suclasses. In classical like walk, we see two subclasses that are corresponding to Gaussian distributions with different Gaussian parameters (see \ref{Classic}). In addition, we matched our results in semi-classical/quantum like walk with those in the presence of decoherence (see \ref{Semi}). These were some of the issues that yet to be addressed in literature. That being said, we see that for state engineering and preparation purposes, walks with a step-dependent coin has high capability. In following sections, we give more details and comparison in the context of entropy.    
\section{Experimental implementation}  \label{Experimental}

The next issue that we should address is the feasibility of an
experimental implementation for SDC walk. The experimental
implementation that we have in mind is inspired by an earlier work
of Schreiber et al \cite{Schreiber}. The setup employs passive
optical elements in order to preform the walk. The walker is a
photon and internal degrees of freedom are its polarizations.
The coin operator could be realized by a half-wave plate changing the polarization of the
photon. The reason why such experimental setup is of interest for
us is due to its flexibility in its coin operator. The matrix
representation of the coin operator changing the polarization of
the photon is given by \cite{Schreiber}
\begin{equation}
C^{\prime }=\left(
\begin{array}{cc}
\cos \left( 2\eta \right) & \sin \left( 2\eta \right) \\
\sin \left( 2\eta \right) & -\cos \left( 2\eta \right)%
\end{array}%
\right) ,
\end{equation}
in which $\eta $ is the rotation angle of the half-wave plate
relative to one of its optical axes. Considering that the coin
operator in the QW introduced in this paper is modified at each
step, the setup of coin in this experimental scheme is ideal one.

The initial state of the walker could be provided by a combination
of a half- and quarter-wave plates. The coin operator is half-wave
plate and step operation is given by an optical feedback loop. A
pulsed laser source provides the photon's wave packet while these
pulses are attenuated to the photon's level by using density
filters. The internal degrees of freedom of the photon are
\emph{Horizontal} and \emph{Vertical} polarizations. In each step,
the horizontal and vertical components of the polarization are
separated spatially and temporally, then recombined and is
sent back to the input beam splitter for the next step. In the
next step, the application of coin operator creates a
superposition of states which their analyze in the internal degree
of the photon (Horizontal and Vertical polarizations) displays
interference. Each step of the walk in this setup corresponds to
one loop. At each step, the photon could be coupled out of the
loop with $50$ percent probability. If such case takes place, an
avalanche photodiode registers a click. This click is recorded by
a computer via a time-to-digital converter interface and
combination of these clicks which are due to a series of
consecutive runs of the experiment, characterizes the walk. Among
the important properties of this experimental setup, one can point
out: I) A high degree of coherence and the scalability of system.
II) keeping the amount of required resources constant as the walker's
position Hilbert space is increased. III) The flexibility of setup
which was essential for the walk introduced in this paper. It is
worthwhile to mention that another experimental setup was also
proposed with similar ingredients by Broome et al in which a
tunable decoherence was added into the system \cite{Broome}.

\section{Shannon Entropy} \label{Shannon Entropy}

Of the fundamental concepts in quantum information and computing
is the Shannon entropy \cite{Nielsen}. The Shannon entropy was
first introduced in 1948 by Claude Shannon \cite{Shannon} in order
to measures the amount of uncertainty that is present in the state
of a physical system. The Shannon entropy has been employed to
address different issues in physical systems such as processing
and transmitting quantum information \cite{Leff}, atomic physics
\cite{Gonzalez-Ferez1}, crystallography \cite
{Menendez-Velazquez}, ultracold trapped interacting bosons
\cite{Haldar} and quantifying the information encoded in complex
network structures \cite{Anand}. In the context of QW, the
Shannon entropy was computed and numerically presented for
different types of generalized Hadamard coin in Refs.
\cite{Bracken,Chandrashekar}. In addition, limit theorems for the
Shannon entropy and its value were computed by using a path
counting method in Ref. \cite{Ide}.

Here, we would like to address the Shannon entropy associated to SDC walk and compare it with SIC case. In the context of QW, the Shannon entropy is calculated using
probability densities of the positions/internal degrees of freedom that are occupied by wave function of the walker in each step. Therefore, the entropy is step dependent. Also, one can 
introduce two different entropies for QW;  I) $S_{C}$, associated to internal degrees of freedom or simply coin space. II) $S_{P}$, related to position space. The Shannon entropy for these two subspaces is given by \cite{Nielsen,Bracken,Ide}
\begin{eqnarray}
S_{P} &=& -\sum_{n}P_{n}LogP_{n},  \label{entrop1}
\\
S_{C} &=& -\sum_{i}P_{i}LogP_{i},      \label{entrop2}
\end{eqnarray}
in which $P_{n}$ and $P_{i}$ are the probability density of the position $n$ and internal degrees of freedom $i$, respectively. The logarithm is given in base of $e$.
According to \eqref{entrop1} and \eqref{entrop2}, for fully localized walk, the entropy is zero
(which is physically expected). Based on previous section, the entropy as a function of step is given for SDC (blue colored lines) and SIC (red colored lines) walks for different rotation
angles in Fig. \ref{Fig3} --q.

For the entropy associated to probability density distribution in position space ($S_{P}$), the following could be stated: 

\textbf{I)} In SIC case, the entropy is an increasing function of the steps, irrespective of
considered $\theta$. Whereas, for the SDC case, the general behavior of entropy depends on $\theta$. 

\textbf{II)} In general, the entropy of SDC walk is
smaller than its corresponding SIC one. This is even valid for the $\theta=\pi/3$ where the
walker is faster and its variance is bigger comparing to its SIC
counterpart (see Fig. \ref{Fig3} --q). The only exception is $\theta=2\pi/5$ (see Fig. \ref{Fig3} --15). This is because, for this coin, SDC walk has a more symmetrical probability density distribution in position space comparing to its SIC counterpart. It is a well established concept that system with more symmetrical probability density distribution has higher entropy \cite{Coles}. 

\textbf{III)}The periodic behavior and localization in probability density distribution of SDC walk could also be detected in entropy's evolution. The periodic behavior is where identical entropy is found and repeated on a regular basis. Localization happens where entropy is zero.

\textbf{IV)} The evolution of entropy for SDC walk shows that: although in each step the number of positions occupied by wave function of the walker increases, this does not necessary mean that entropy is also increasing (in contrast to SIC walk). In fact, for certain steps, minimums are formed for the entropy.

In case of the entropy associated to probability density distribution in coin space ($S_{C}$), one can highlight the following points: 

\textbf{I)} Since the coin space contains two elements (two internal degrees of freedom),  entropy of the coin space is bounded by an upper limit. The highest value of $S_{C}$ is obtained when probability density distribution in coin space is equally distributed over these two elements. Therefore, the maximum $S_{C}$ that walker with two internal degrees of freedom can obtain is $0.693147$.

\textbf{II)} For walks with SIC, a damped oscillation is observed for $S_{C}$. This shows that by increasing the number of steps, the entropy of coin space converges to a specific value. For SDC case, the general behavior depends on the $\theta$. For semi-classical/quantum and quantum like walks, a damped oscillation is also seen for the $S_{C}$ (see lower panels of Figs. \ref{Fig3} --15 and \ref{Fig3} --17). In contrast, for localized and compact classical like walks, the entropy of coin space has periodic behavior. Finally, for the classical like walk, it has an irregular behavior but converging one.  

\textbf{III)} The complete localization for the walker takes place when both the entropy of position and coin spaces are zero. Such cases take place for specific steps in localized and compact classical like walks (see Figs. \ref{Fig3} --12 and \ref{Fig3} --13). 

\begin{figure*}[!htbp]
 {\begin{tabular}[b]{ccc}%
    	\renewcommand*\thesubfigure{\arabic{subfigure}}	
    	\setcounter{subfigure}{0}		
    	\subfloat[$\theta=\pi/4$]{\includegraphics[width=0.3\textwidth]{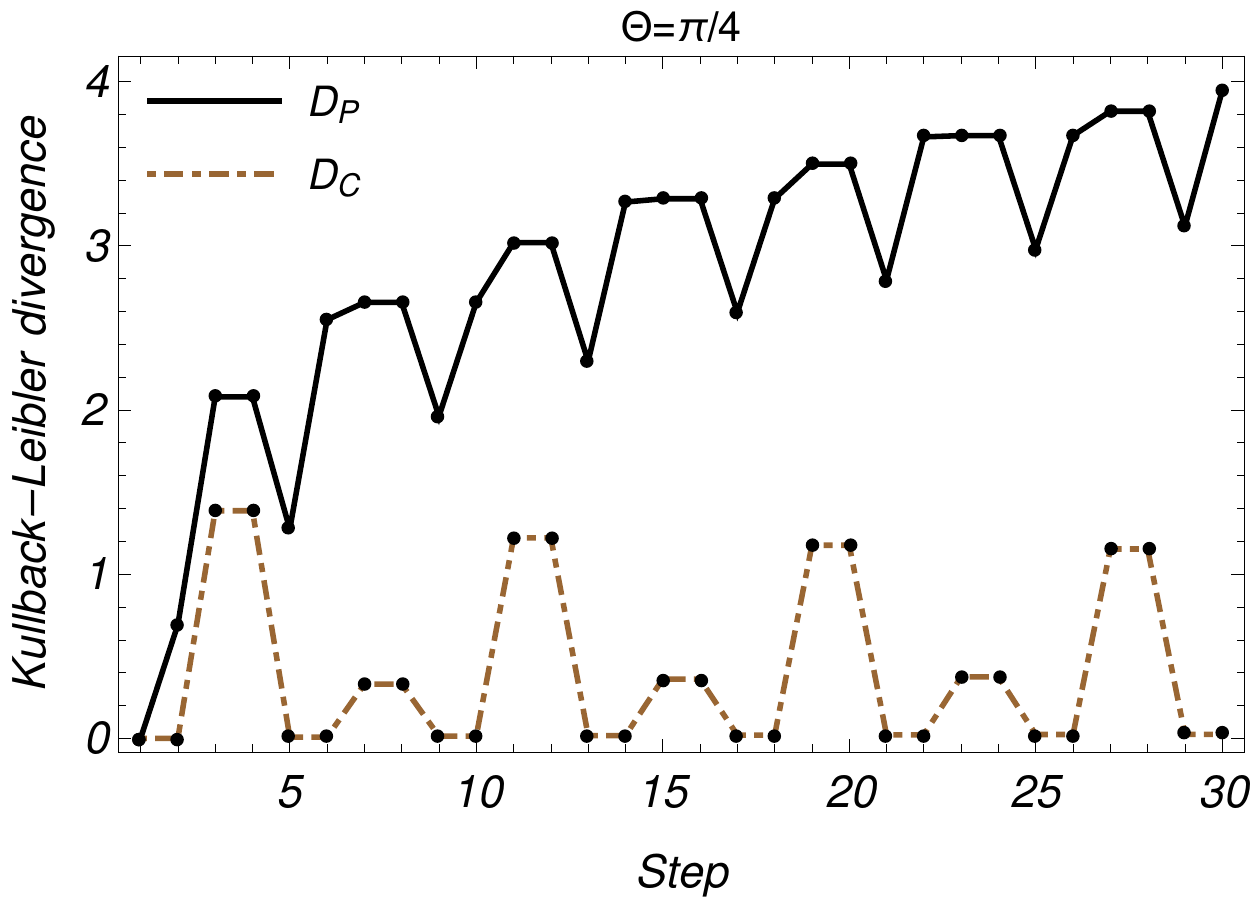}}  	
    	\subfloat[$\theta=\pi/12$]{\includegraphics[width=0.3\textwidth]{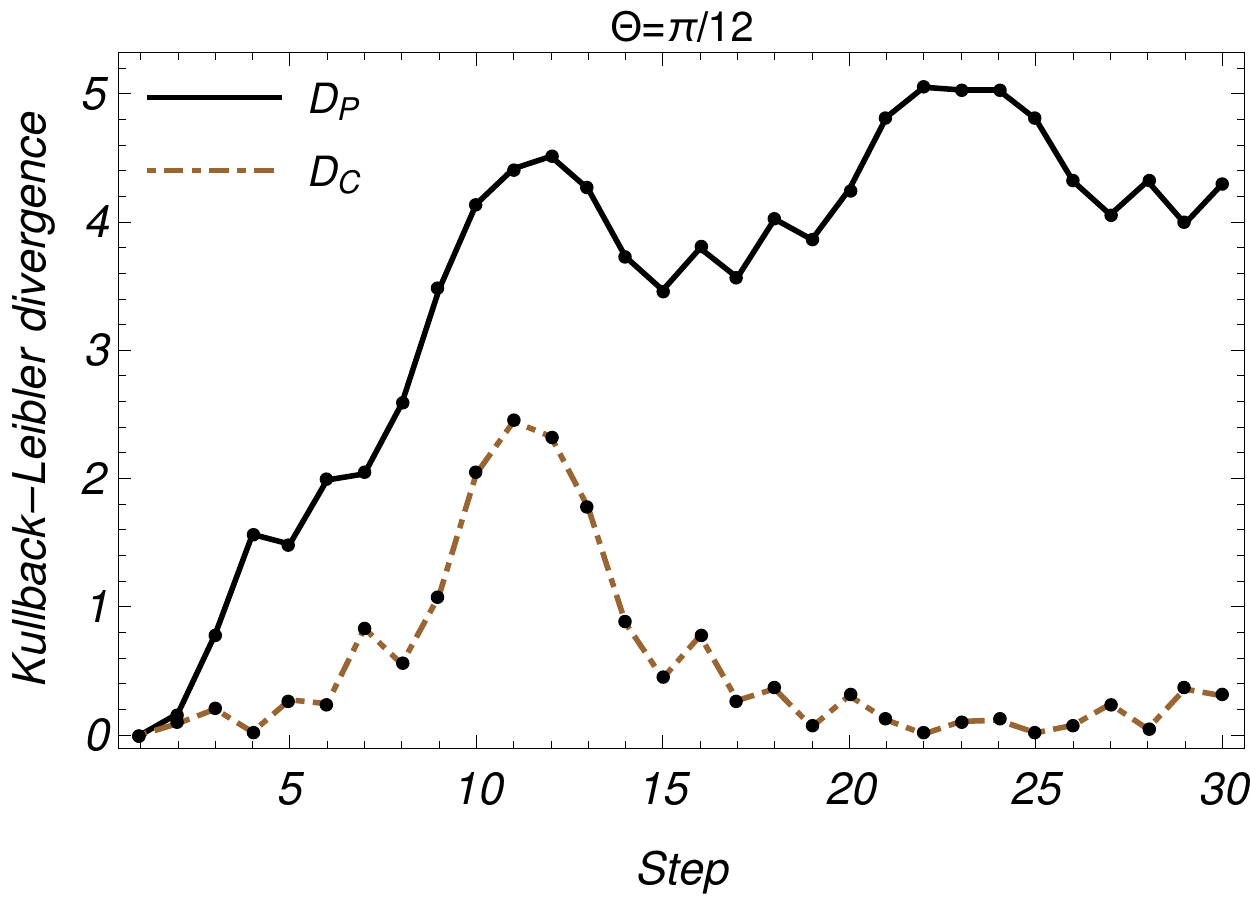}} 
    	\subfloat[$\theta=3.59\pi /5$]{\includegraphics[width=0.3\textwidth]{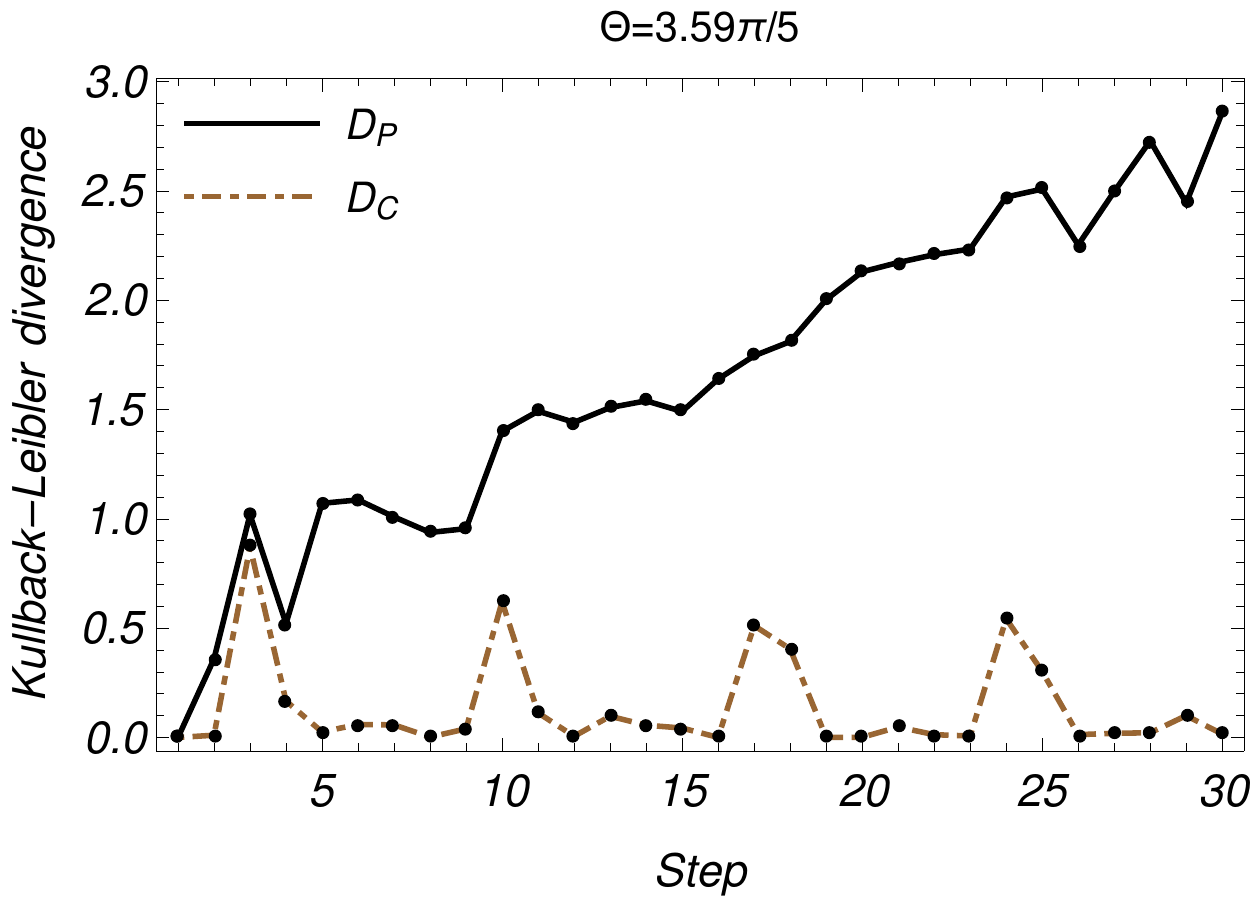}}  \\[0.01cm]
    	\renewcommand*\thesubfigure{\arabic{subfigure}}	
    	\subfloat[$\theta=2\pi/5$]{\includegraphics[width=0.3\textwidth]{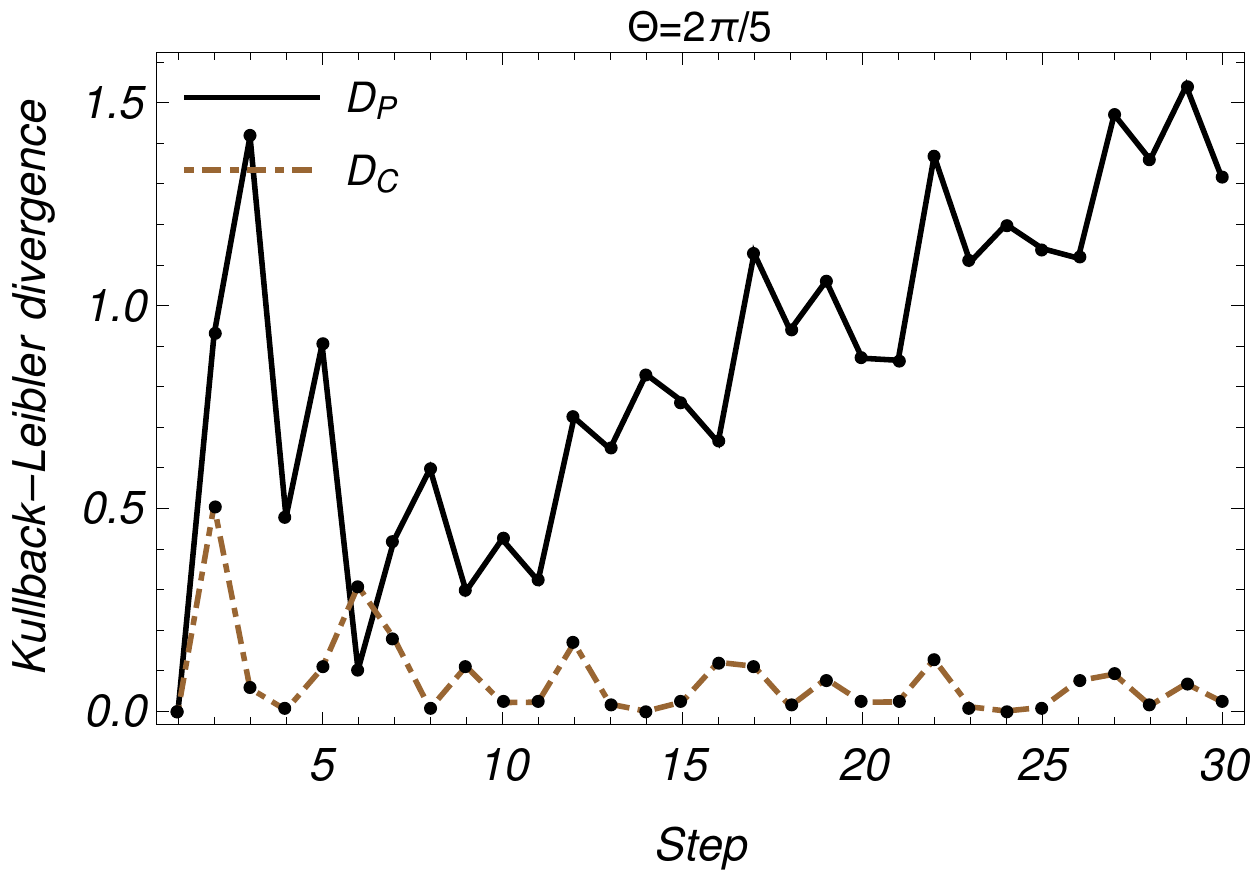}} 	
    	\subfloat[$\theta=\pi/5$]{\includegraphics[width=0.3\textwidth]{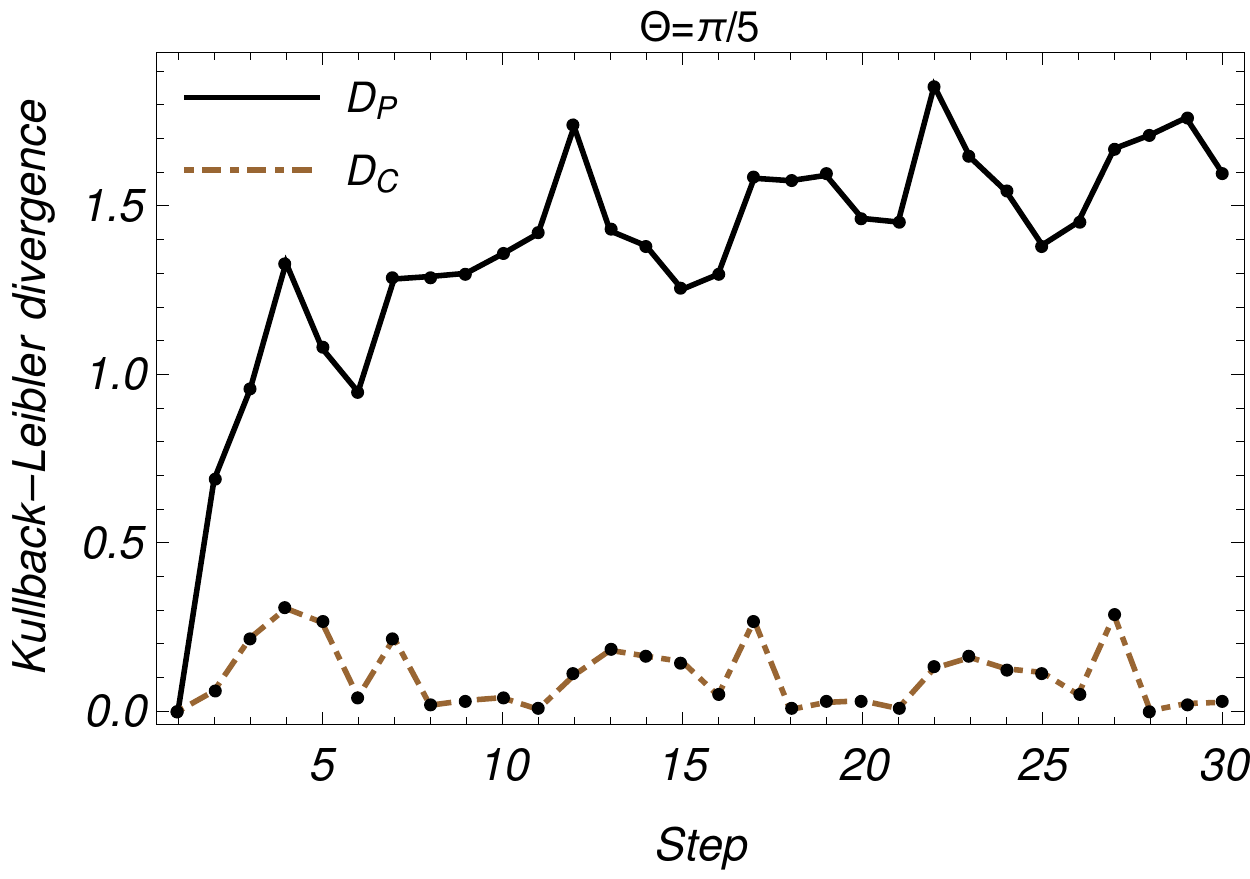}}  
    	\subfloat[$\theta=\pi/3$]{\includegraphics[width=0.3\textwidth]{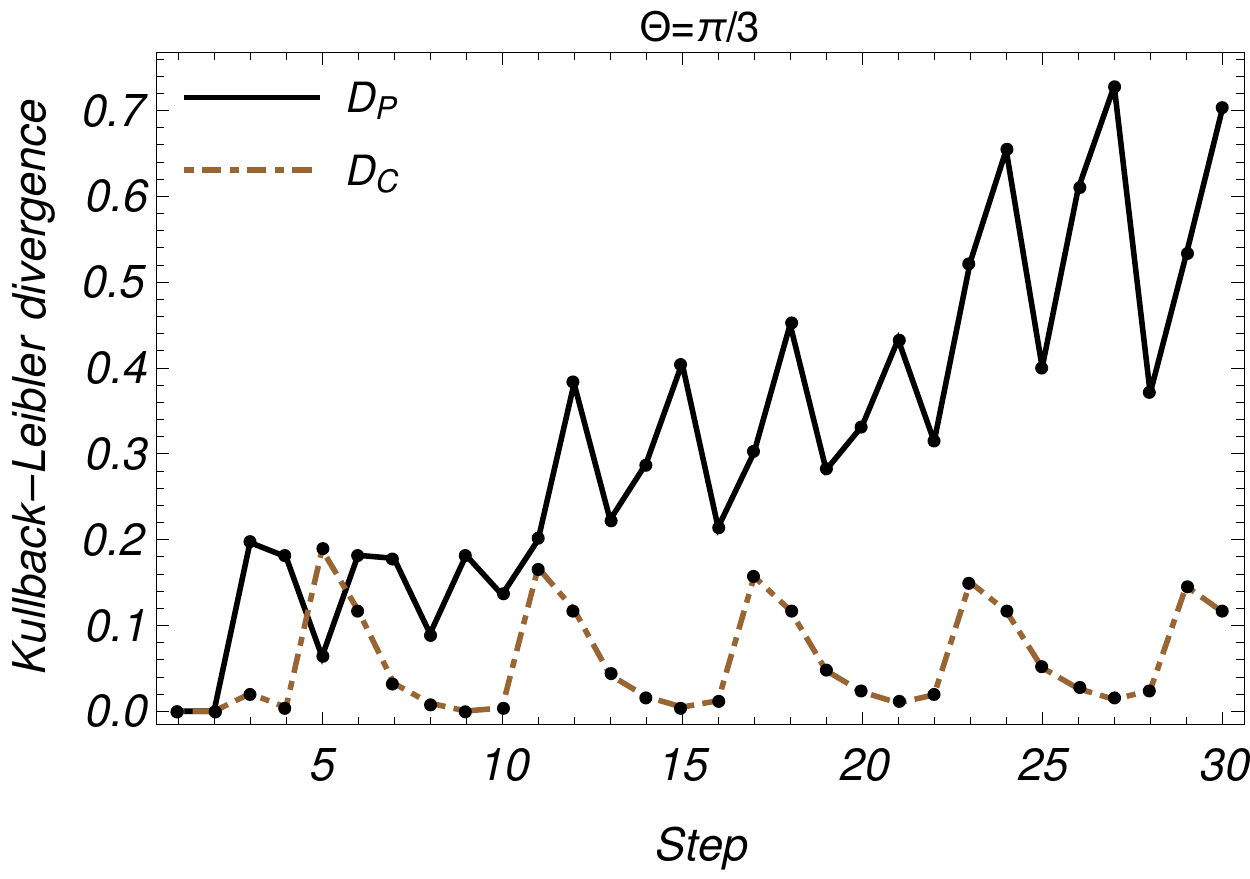}}  
    \end{tabular}}   \\
    \caption{Kullback-Leibler divergence between SDC and SIC walks for different rotation angles. } \label{FigII}
\end{figure*}

In order to understand the differences between entropy of SDC and SIC walks in more details, we investigate the relative entropy. To do this, we use Kullback-Leibler divergence which measures how one probability density distribution diverges from the other one. The Kullback-Leibler divergence is given by 
\begin{eqnarray}
D_{P} &=& \sum_{n}P_{n}Log\frac{P_{n}}{P'_{n}},  \label{entrop3}
\\
D_{C} &=& \sum_{i}P_{i}Log\frac{P_{i}}{P'_{i}},      \label{entrop4}
\end{eqnarray}
where $P'_{n}$ and $P'_{i}$ are the probability density of the position $n$ and internal degrees of freedom $i$ for SIC walk, respectively. Kullback-Leibler divergence is zero when the probability density distributions for SDC and SIC walks are identical. The results for different rotation angles are presented in Fig. \ref{FigII}.
For Kullback-Leibler divergence of position space, $D_{P}$: In general, we have the following relation for Kullback-Leibler divergence of different rotation angles

\begin{eqnarray*}
D_{P}(\theta=\frac{\pi}{3}) &<& D_{P}(\theta=\frac{2\pi}{5})<D_{P}(\theta=\frac{\pi}{5})< \\
D_{P}(\theta=\frac{3.59\pi}{5})&<& D_{P}(\theta=\frac{\pi}{12})<D_{P}(\theta=\frac{\pi}{4}). 
\end{eqnarray*}  

This relation reminds us the earlier classification for different rotation angles. We see that in this context as well, our classification could be applied. The smallest Kullback-Leibler divergence between SDC walk and its SIC counterpart is observed for the quantum like walk ($\theta=\pi/3$) while the largest one is found for localized walk ($\theta=\pi/4$). Overall, the Kullback-Leibler divergence between SDC and SIC walks increases but its rate of increment depends on $\theta$. 

For Kullback-Leibler divergence of coin space, $D_{C}$: In some steps, $D_{C}$ becomes zero indicating that probability density distributions in coin space for SDC and SIC walks are identical. In general, some semi-periodic behavior is observed for different rotation angle. The only exception is for $\theta=\pi/12$. Overall, we notice that $D_{C}$ tends to decrease indicating that for probability density distributions in coin space for SDC and SIC walks become more similar to each other. 

The Shannon entropy is one of the central concepts in quantum
information theory. According to Nielsen and Chuang
\cite{Nielsen}, this quantity measures the amount of uncertainty
(information/randomness) that is present in the state of a physical system.
Therefore, smaller entropy indicates lower uncertainty
(information/randomness). Through our calculations, it was shown that SDC
walk enjoys smaller entropy of the position space comparing to its SIC counterpart (with
some exceptions). Therefore, the SDC imposes smaller uncertainty
in the state of walker comparing to SIC. One can also conclude
that amount of information (randomness) acquired/modified through SDC evolution
is smaller than the one obtained/modified through SIC evolution.
In the context of data communication, entropy provides the limit
on length that a lossless compression encoding of the data could
be achieved. If the entropy of source is smaller than the capacity
of communication channel, the receiver could reliably attains the
communication from the source. Therefore, we see that SDC case
provides an entropy which widens the limiting range and includes
larger possibilities in the context of data communication
(comparing to SIC walk). In addition, we should once more emphasize on
the fact that despite expectation, the increase in number of the
positions occupied by wave function of the walker does not
necessarily mean increase in entropy. The entropy measures
disorder in the system. Generally, the common sense indicates that
if the wave function of the walker is dispersed over larger number
of the positions between two steps, the entropy should increase.
But as we saw for SDC case, this is not the case. In fact, the
wider dispersion over position resulted into smaller entropy for
specific steps. 

\section{Conclusion} \label{Conclusion}

In this paper, we studied a class of quantum random walk with a
step-dependent coin. Such a coin can result in a reach behavior of
the walker. It was shown, in particular, that such a coin may
result, for different choices of the rotation angle, into four
distinguishable behaviors in probability density of walker
including: complete localization, (compact) classical like,
semi-classical/quantum like and quantum like. This diversity in probability density distribution signals the controllability over QW where controlling factor is rotation angle. In the context of
speed, it was shown that the walk with a step-dependent coin could
be faster (slower) comparing to its step independent counterpart.

In addition, we studied the entropy of walks with the step-dependent coin and compare it with its corresponding step-independent coin case. The entropy was calculated for two cases of probability density distributions in coin ($S_{C}$) and position ($S_{P}$) spaces. Interestingly, it was shown that for the
walk with step-dependent coin, the $S_{P}$ is smaller than the
walk with step-independent coin (with some exceptions). In
addition, it was shown that general behavior of the $S_{P}$ as a
function of steps, depends on choices for the rotation angle in walk
with a step-dependent coin. Whereas for walks with the step-independent coin, irrespective of choices for rotation angle, a monotonic
behavior was observed for $S_{P}$ as a function of steps. Finally,
it was pointed out that by increasing the number of positions
occupied by wave function of the walker, $S_{P}$ does not increase. 
In fact, for special step-dependent coins, $S_{P}$ could decrease. Regarding to the $S_{C}$, 
a damped oscillation behavior was observed for walks with step-independent coin. In the contrast, for step-dependent coin walk, three distinctive behaviors were observed for $S_{C}$ as a function of steps: a damped oscillation, periodic and irregular behaviors. 

The study conducted in this paper can be generalized also to
multi-walker systems, higher dimensional walk as well as to
(decoherent) walks that couple to their environment.  These QWs
can be applied to explore topological phases
\cite{Kitagawa,Cardano} and especially \textit{new} classes of
such topological phases. In addition, it would be interesting to
see how different entropy properties that were obtained for the
walker here could be employed in the context of data compression,
communication, and cryptography. We will leave these matters for
future works.

\section{Appendix A: Justification for classification}

Through out the paper, it was shown that using step-dependent coin results into obtaining diverse probability density distribution in position space. We classified these distributions according to their properties. Among them, we had the classical like and semi-classical/quantum like behaviors.

\subsection{Classical like behavior} \label{Classic}

\begin{figure}[!htbp]
    \centering
  	\subfloat[Continuous line: SDC walk with $\theta=\pi/12$. Dashed line: Gaussian distrbiution with $\sigma=0.5$ and $\mu=5$.]{\label{C1}\includegraphics[width=0.3\textwidth]{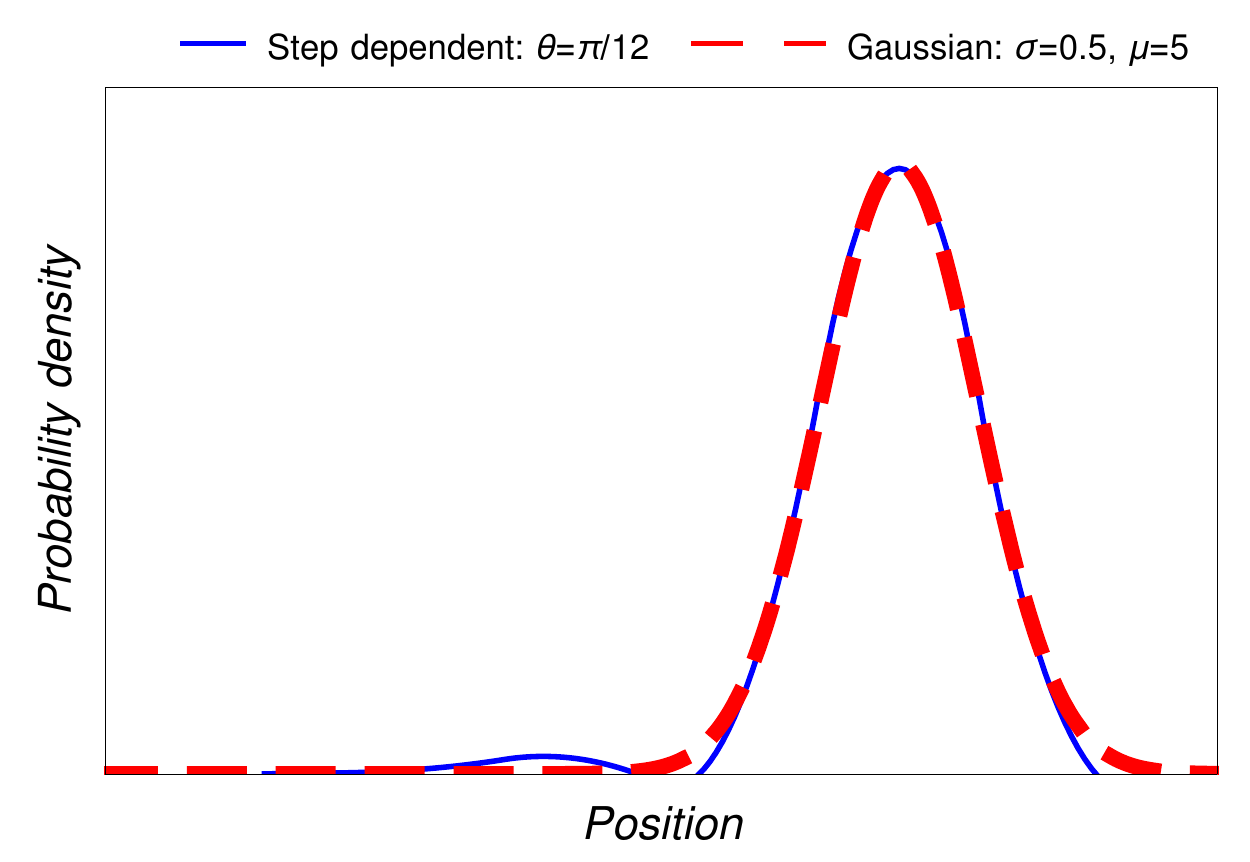}}
  	\quad
  	\subfloat[Continuous line: SDC walk with $\theta=3.58\pi/5$. Dashed line: Gaussian distrbiution with $\sigma=0.61$ and $\mu=3.39$.]{\label{C2}\includegraphics[width=0.32\textwidth]{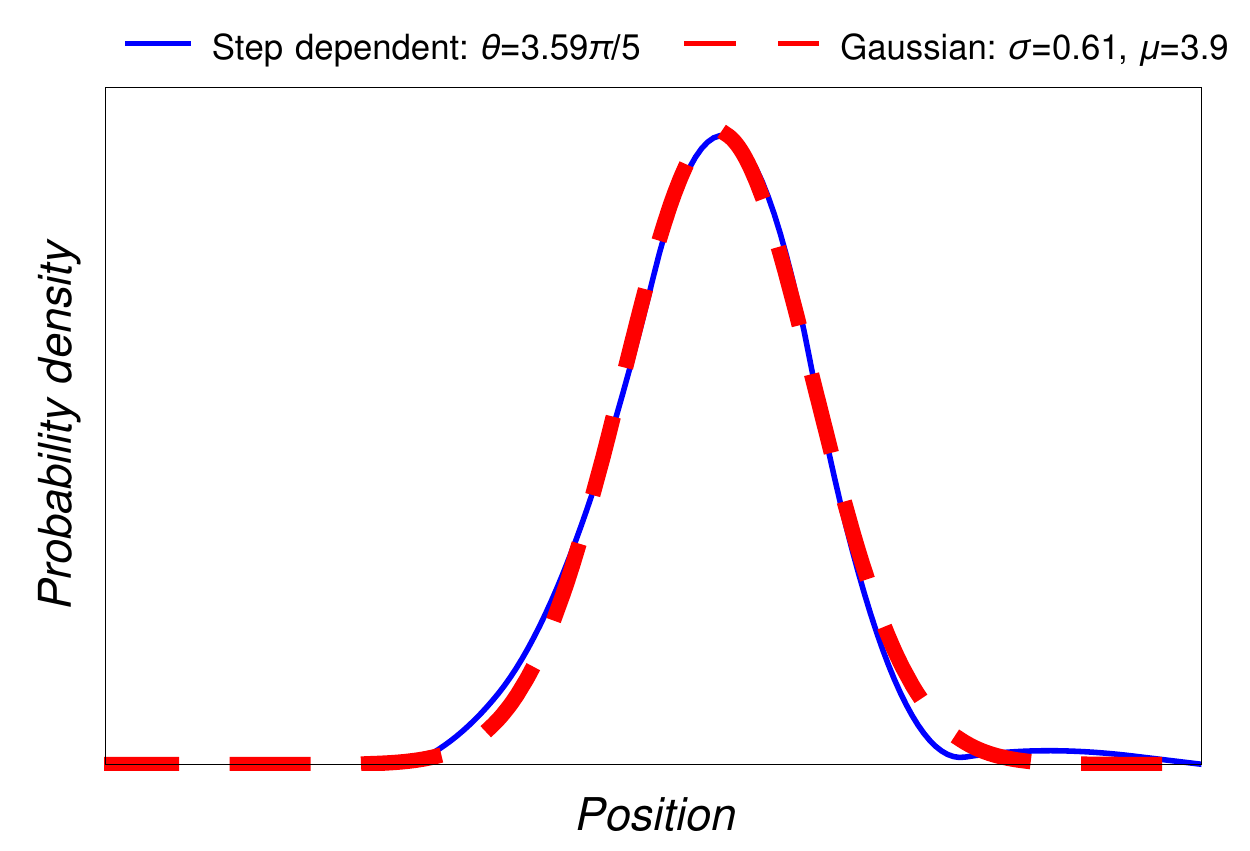}}
    \caption{Probability density distribution for $6th$ step.}
    \label{Fig4}
  \end{figure}

The classical behaviors are the ones that present normal (Gaussian) like distributions for their probability density distributions. The probability density of the normal distribution is given by 
\begin{equation}
f(x\mid \mu, \sigma^2)= \frac{e^{-\frac{(x-\mu)^2}{2\sigma^2}}}{\sqrt{2 \pi \sigma^2}},
\end{equation}
where $\mu$ is the mean or expectation of the distribution, $\sigma$ is the standard deviation, and $\sigma^2$ is the variance. In our study, we had two subclasses for classical like behaviors for different rotation angles: compact and usual classical likes. To show that mentioned cases actually have Gaussian like distribution, we study the $6th$ step of these walks and compare their probability density distributions with Gaussian ones. The task is to estimate the parameters of the Gaussian distribution that would correspond to the mentioned cases. This could be done by using the number of positions occupied by the wave function and their probability density. The results are given in following diagrams (see Fig. \ref{Fig4}).

Evidently, by tuning the Gaussian parameters, the probability density distributions of the SDC QW with considered rotation angles could be highly fitted with those of Gaussian distributions. The mean variance considered for fitting the Gaussian distributions show that one of these distributions is indeed more compact comparing to the other one. These facts justify our earlier classification of the walks with these coins into classical like behavior and their corresponding subclasses. Although this is done for the $6th$ step, the same could be done for other steps as well. It should be noted that in each step, the Gaussian parameters would be different.

\subsection{Semi-classical/quantum like} \label{Semi}

\begin{figure*}[!htbp]
	{\begin{tabular}[b]{ccc}%
	\renewcommand*\thesubfigure{\arabic{subfigure}}	
	\setcounter{subfigure}{0}		
	\subfloat[$\ketm{\phi}_{int}=\ketm{0}_{C} \otimes \ketm{0}_{P}$ and $q=0.8$; $F(P,Q)=0.973$.]{\includegraphics[width=0.3\textwidth]{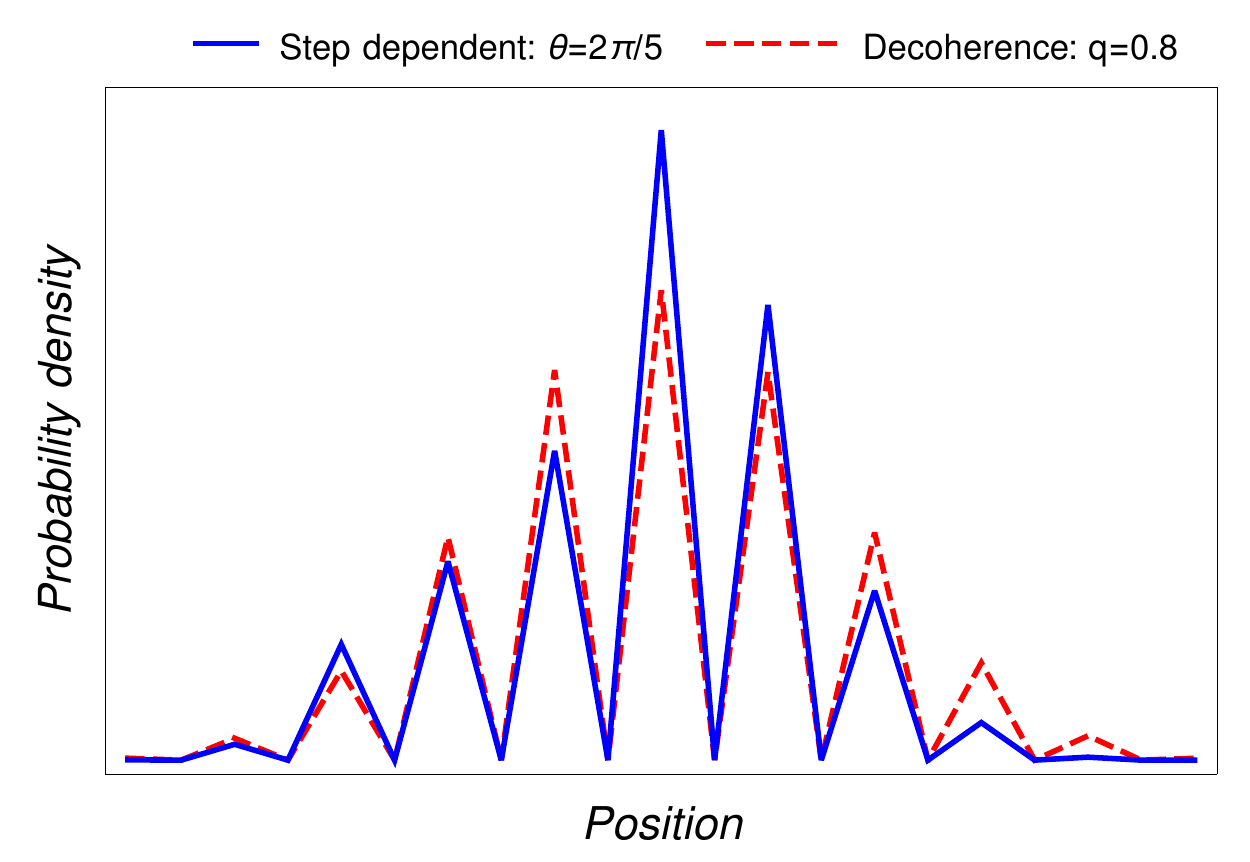}} \label{E2}  	
	\subfloat[$\ketm{\phi}_{int}=(\ketm{0}_{C} +I\ketm{1}_{C} ) \otimes \frac{\ketm{0}_{P}}{\sqrt{2}}$ and $q=0.8$; $F(P,Q)=0.987$.]{\includegraphics[width=0.3\textwidth]{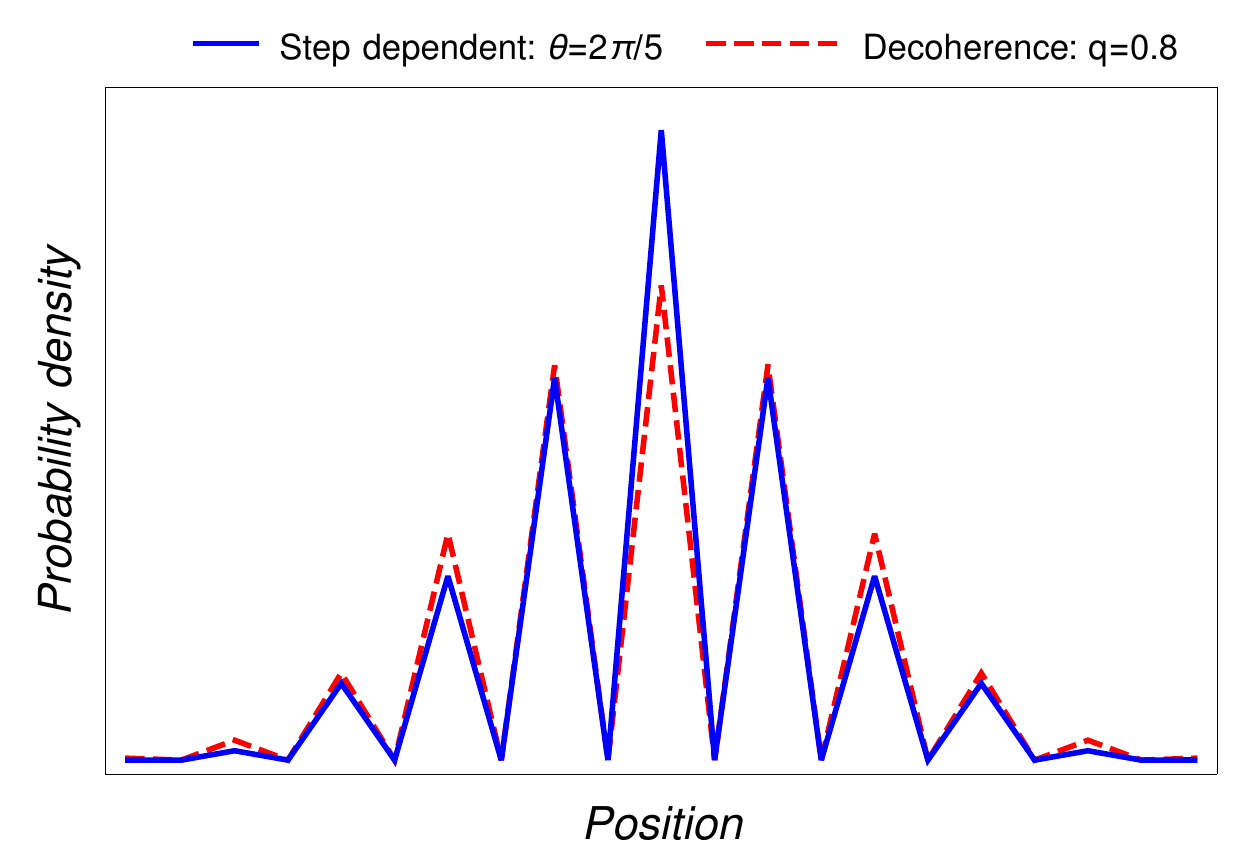}} \label{E5}
	\subfloat[$\ketm{\phi}_{int}=\ketm{1}_{C} \otimes \ketm{0}_{P}$ and $q=0.8$; $F(P,Q)=0.973$.]{\includegraphics[width=0.3\textwidth]{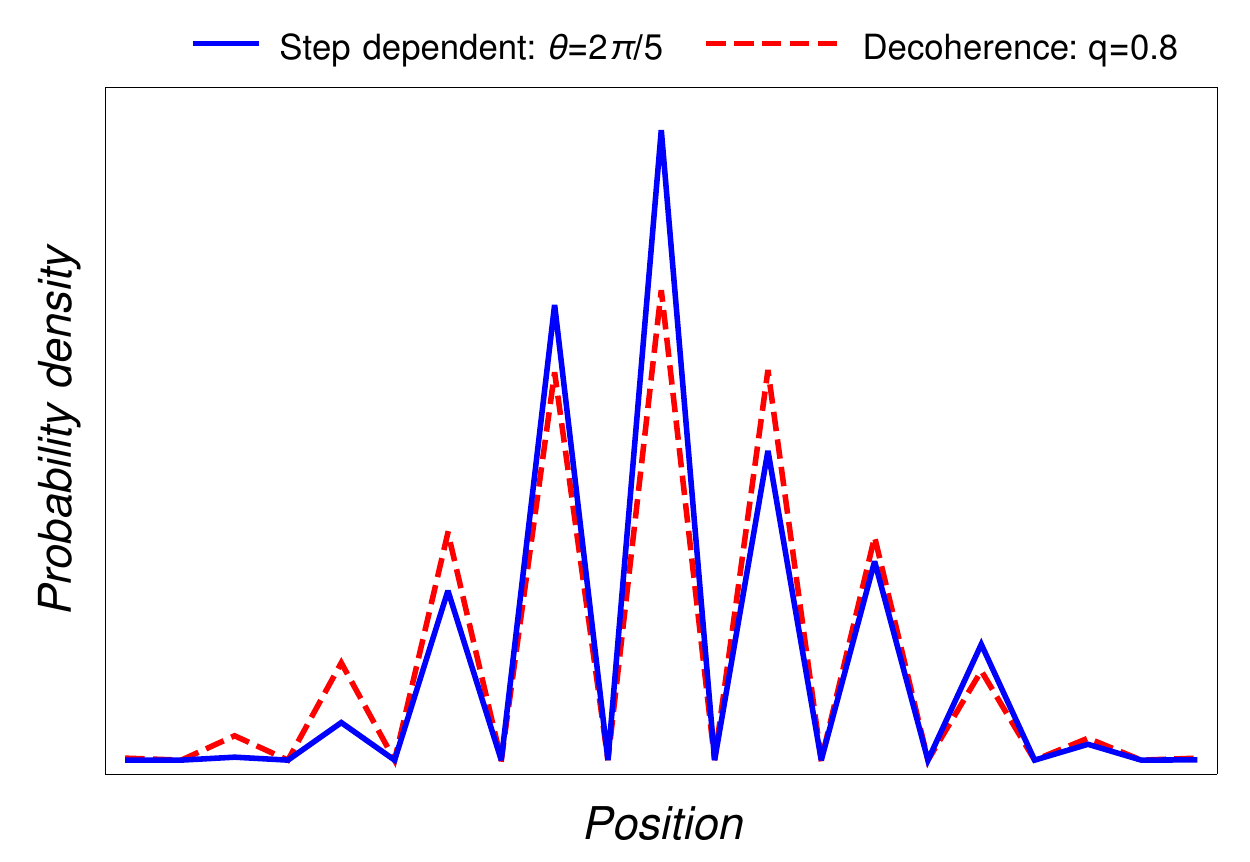}} \label{E6} 
	\end{tabular}}   \\
	\caption{Probability density distribution for $10th$ step; Continuous line: SDC walk with $\theta=2\pi/5$. Dashed line: decoherence walk with Hadamard coin and $q=0.8$.} \label{Fig5}
\end{figure*}

The next classification that we address is semi-classical/quantum like. To do so, we consider the walks with decoherence. It is a well established concept that depending on the amount of decoherence in QW, system's behavior could range from purely quantum like to classical one \cite{Brun1,Brun,Alberti1}. There are different methods to include decoherence in QW. One of them is by measuring the coin and/or position space after each step. Inclusion of decoherence results into losing the unitary nature of the process. Therefore, one should use density operator formalism to address the nonunitary evolution of the walk at each step given by \cite{Brun1,Brun,Alberti1} 
\begin{eqnarray}
\hat{\rho}(t+1) & = &  (1-q-s) \hat{U} \hat{\rho}(t) \hat{U}^\dagger+q \mathbb{P}_{C} \hat{U} \hat{\rho}(t) \hat{U}^\dagger \mathbb{P}_{C}^\dagger \notag \\[0.1cm]
&   & \quad +\:
s \mathbb{P}_{P} \hat{U} \hat{\rho}(t) \hat{U}^\dagger \mathbb{P}_{P}^\dagger ,
\end{eqnarray}
where $\mathbb{P}_{C}$ and $\mathbb{P}_{P}$ are projections to coin and position spaces, respectively. $q$ and $s$ are the probabilities of a decoherence event happening per time step. Here, we restrict our study to $s=0$. The value of decoherence parameter will be within the range of $0\leq q\leq 1$, where $q=0$ and $q=1$ correspond to purely quantum and classical like behaviors, respectively. The coin operator of the decoherence case is Hadamard. For SDC walk, we consider the rotation angle of $\theta=2\pi/5$ which had semi-classical/quantum like behavior. We consider three different initial states. The results are presented in Fig. \ref{Fig5} for the $10th$ step.

Accordingly, we observe that by tunning the decoherence parameter, the results of SDC walk with rotation angle of $\theta=2\pi/5$ and walk with decoherence could be fitted together. To mathematically confirm this issue, we investigate the fidelity for these two probability density distributions given by 
\begin{equation}
F(P,Q)=(\sum_{i} \sqrt{P_{i}Q_{i}})^2,  \label{fid}
\end{equation}
where $P_{i}$ and $Q_{i}$ are the probability density of the position $i$ for SDC and decoherent walks, respectively. The $F(P,Q)$ is bounded by $0\leq F(P,Q)\leq1$ where $F(P,Q)=1$ corresponds to have identical distributions. Our calculations shows that fidelity for these cases is more than $0.97$ (please see Fig. \ref{Fig5}), hence two walks being fitted together. Since $q<1$, the walk with decoherence is semi-classical/quantum like. Therefore, our earlier classification for SDC walk with rotation angle of $\theta=2\pi/5$ is justified. Although this is done for $10th$ step, the same process could be done for other steps as well confirming the validity of our argument. It should be noted that if one considers $s\neq0$, the fitting would even improve further. For further comparison between SDC walk with rotation angle of $\theta=2\pi/5$ and decoherent one, please compare Fig. \ref {G2P5} with Fig. $3$ in Ref. \cite{Alberti1}.

\section{Appendix B}

In this section, we expand the discussion of walker's behavior with SDC for different rotation angles. We will construct the Bloch vectors at each position for subsequent steps. The results are presented in Fig. \ref{Fig6}.

\begin{samepage} 
  \begin{figure*}[!htbp]
	  \renewcommand*\thesubfigure{\arabic{subfigure}}
	  \centering
	  {\begin{tabular}[b]{ccc}
	  \subfloat[$\theta=0$]{\label{BP0}\includegraphics[width=0.34\textwidth]{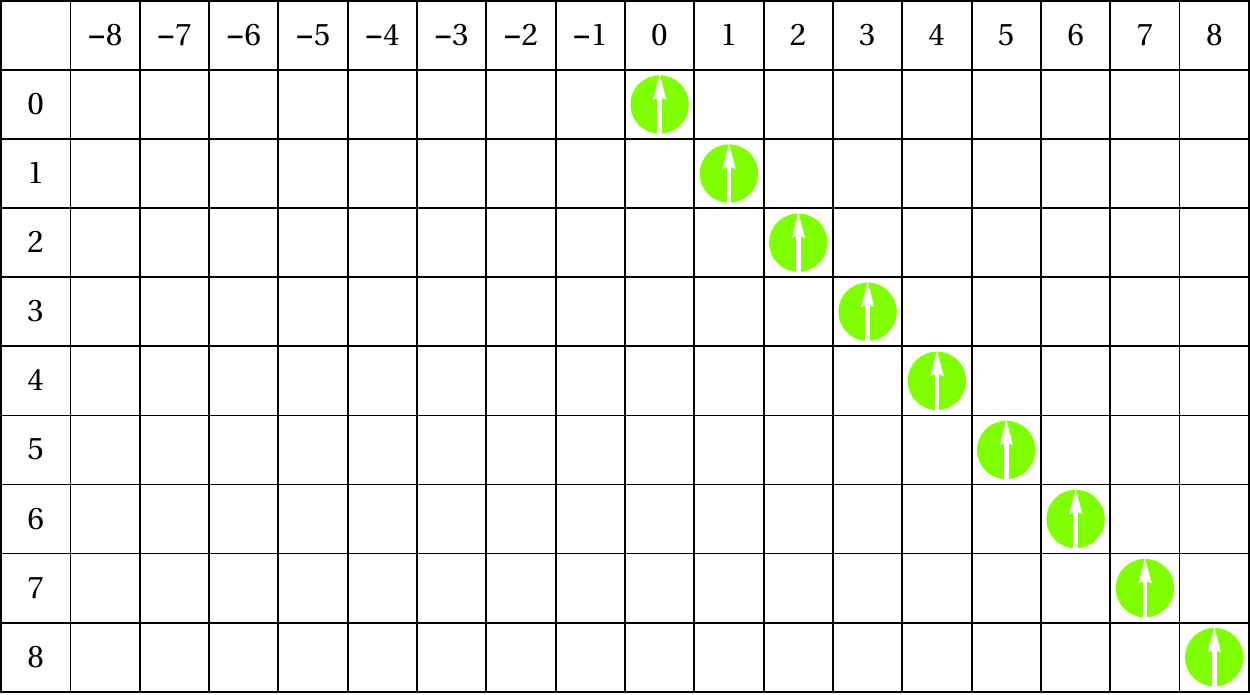}}
	  \quad 
	  \subfloat[$\theta=\pi/2$]{\label{BP2}\includegraphics[width=0.34\textwidth]{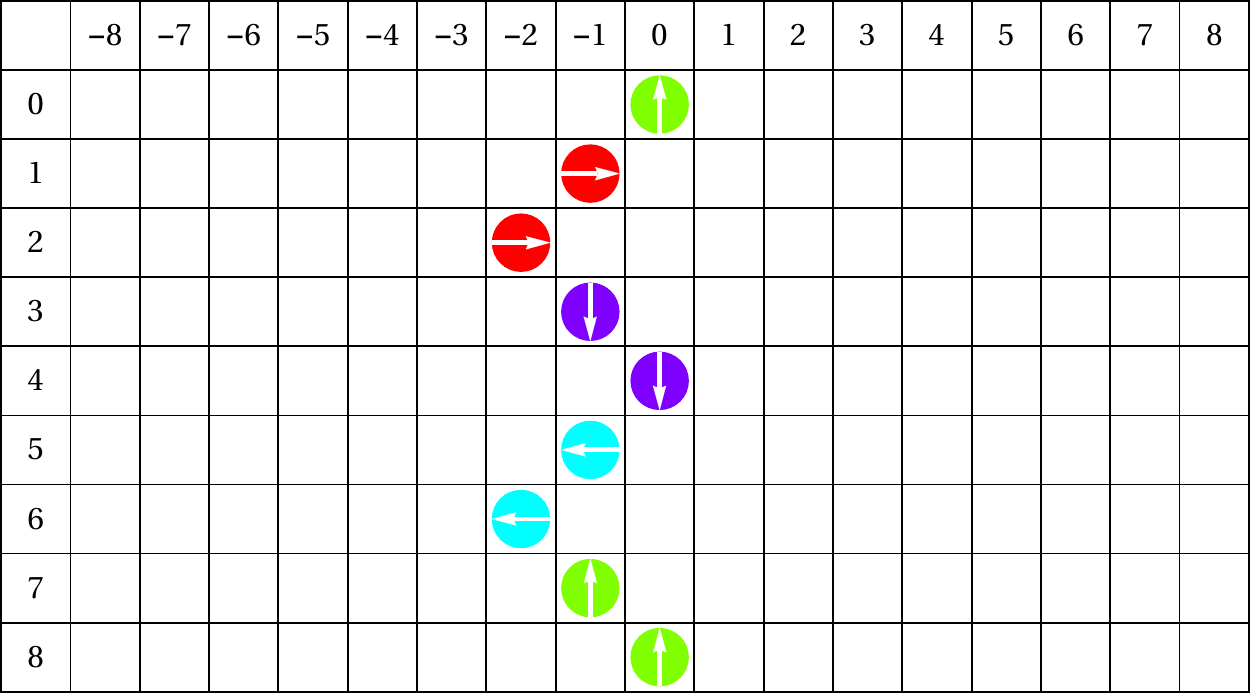}} 	  \quad 
	  \subfloat[$\theta=\pi/4$]{\label{BP4}\includegraphics[width=0.34\textwidth]{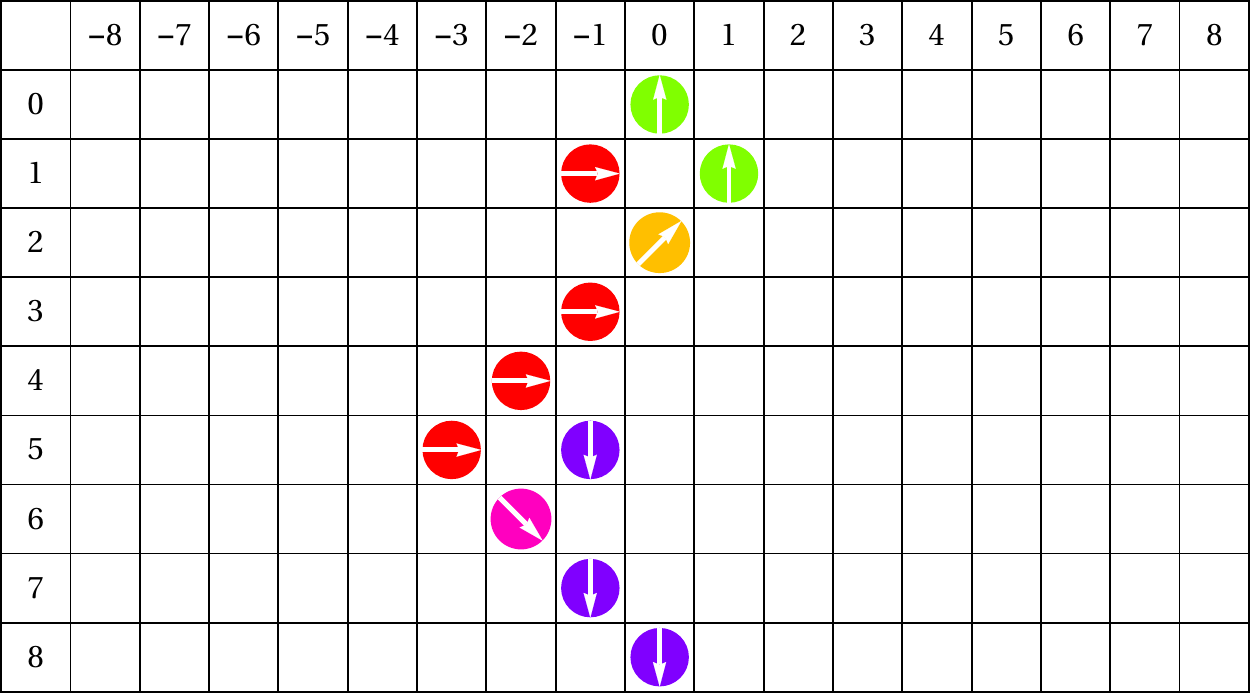}} 	
	  \\  	
	  \subfloat[$\theta=\pi/12$]{\label{BP12}\includegraphics[width=0.34\textwidth]{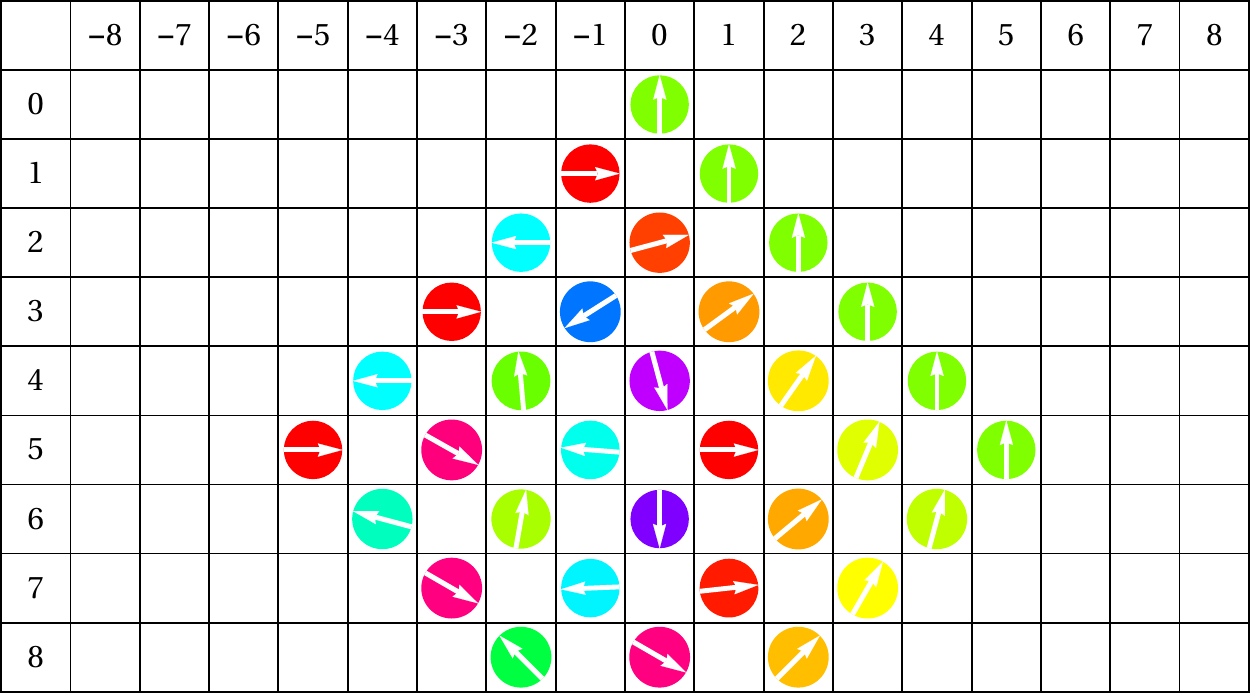}} \quad 
	  \subfloat[$\theta=3.59\pi /5$]{\label{B15P12}\includegraphics[width=0.34\textwidth]{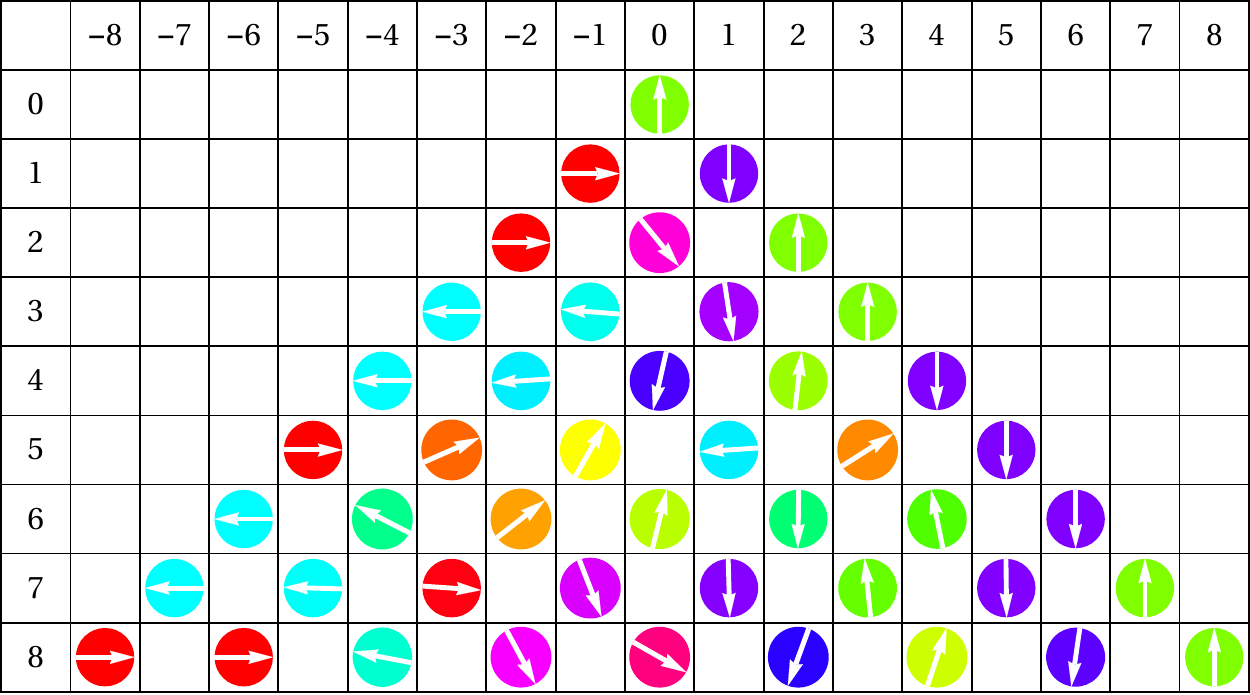}} 
	  \quad 
	  \subfloat[$\theta=2\pi/5$]{\label{B2P5}\includegraphics[width=0.34\textwidth]{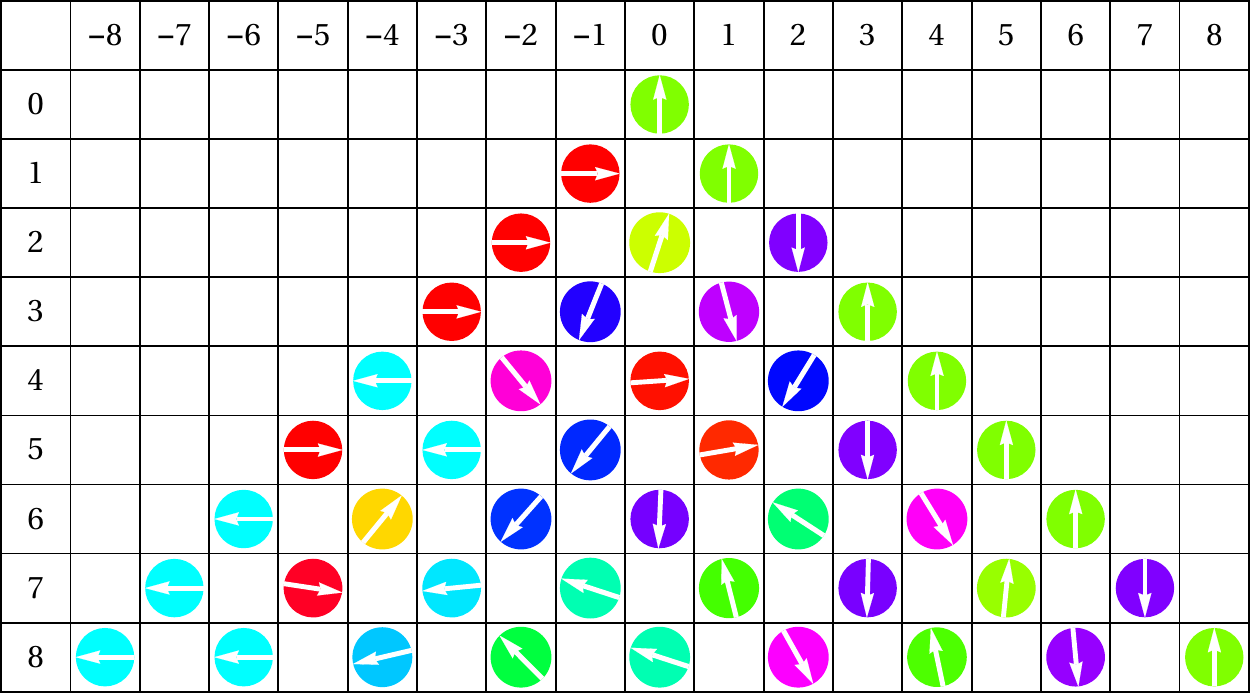}} 
      \\ 
	  \subfloat[$\theta=\pi/5$]{\label{BP5}\includegraphics[width=0.34\textwidth]{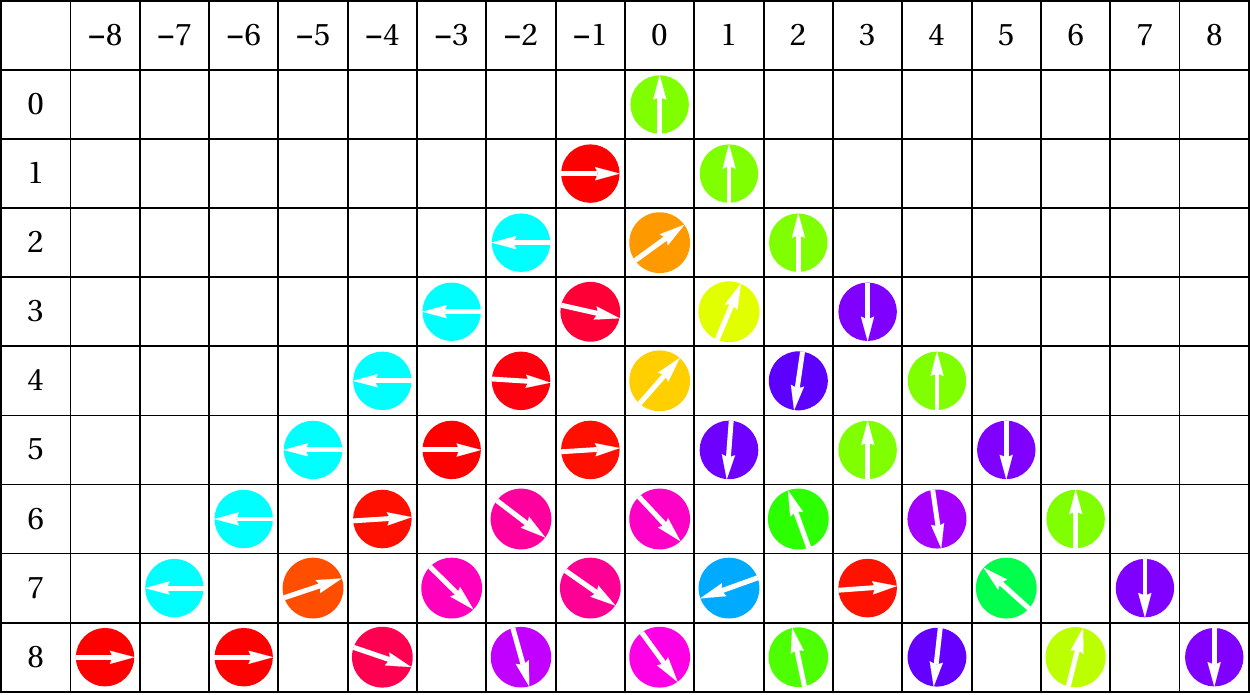}} 	  \quad 
	  \subfloat[$\theta=\pi/3$]{\label{BP3}\includegraphics[width=0.34\textwidth]{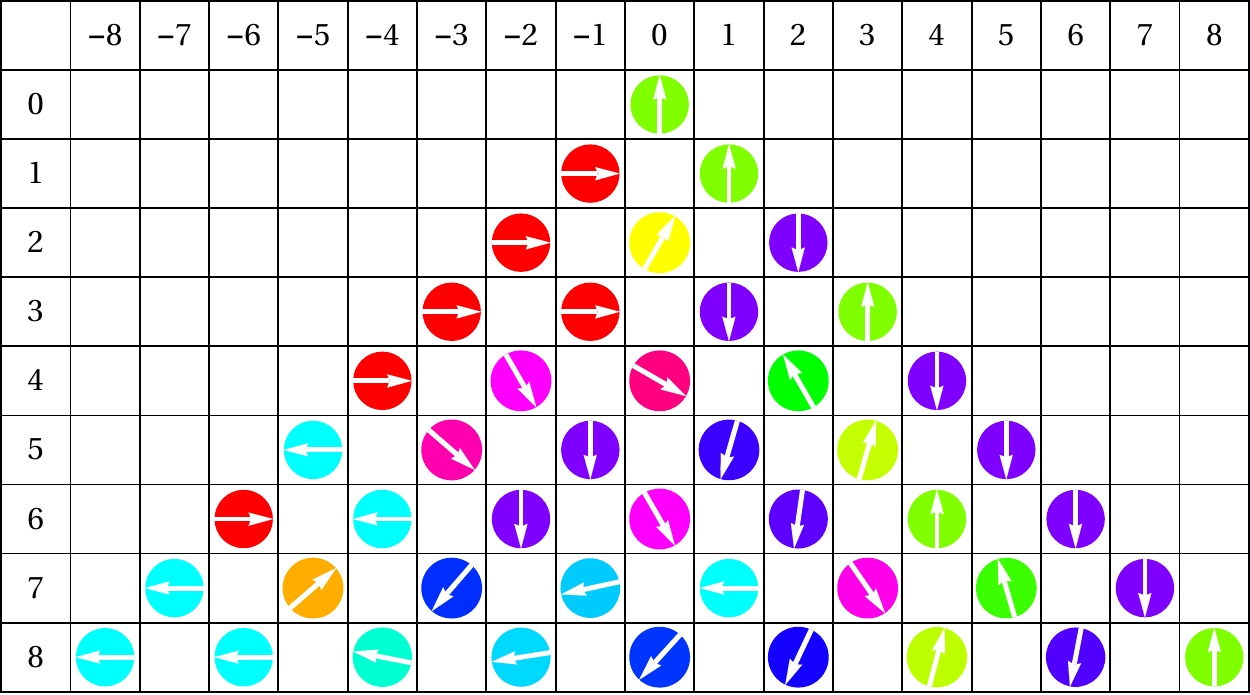}}
	  \quad 
	  \subfloat[For $0.6\ketm{0}_{C}+0.8\ketm{1}_{C}$]{\label{GG}\includegraphics[width=0.2\textwidth]{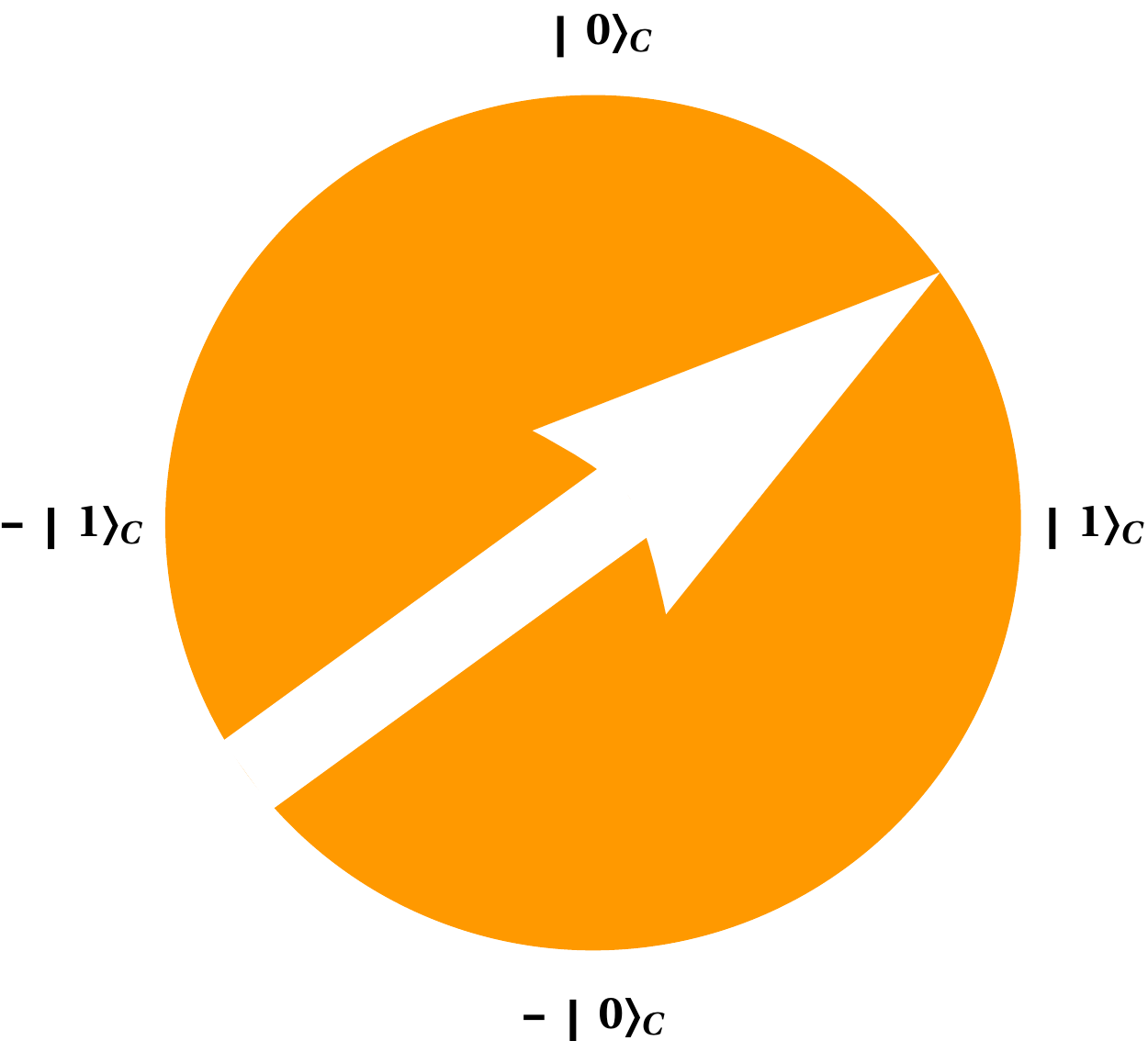}}
	  \end{tabular}} 	
	  \caption{Reconstructed Bloch vectors at each position for eight subsequent steps $T = 0..8$.}
	  \label{Fig6}
   \end{figure*}
For subclasses of localized walk, we the following modifications in internal degrees of freedom of the walker: in free localized walk ($\theta=0$), the distribution over internal degree of the walker remains unchanged at each step (see Fig. \ref{BP0}). In bounded localized walk ($\theta=\pi/2$), in every two steps, the Bloch vectors remains the same and then rotated clockwise with the rotation angle of $\pi/2$ (see Fig. \ref{BP2}). For $\theta=\pi/4$, we have two specific behaviors: I) Distribution equally over two positions; in which the Bloch vectors at one position is perpendicular to the other one. II) Complete localization; where distribution in coin space could be in superposition (steps $3$ and $6$ in Fig. \ref{BP4}). 

For other cases where we have distribution in wave function of the walker, we can highlight two important points: first of all, the Bloch vectors at the leftmost and rightmost hand side positions are always perpendicular to each other (see Figs. \ref{B15P12} -- \ref{BP3}). In these positions, the Bloch vectors could be rotated with rotation angle of $\pi$ or remains the same. The steps where rotation is taking place is different for the leftmost and rightmost hand side positions (depending highly on the coin under consideration). The only exception is for walk with $\theta=\pi/12$. In this case, the wave function of the walker periodically first becomes distributed and then completely localized. When the wave function of the walker becomes distributed, Bloch vectors at the leftmost and rightmost hand side positions are perpendicular to each other. But when the wave function of the walker starts to localize again, this perpendicular property is vanished and at the leftmost and rightmost hand side positions, the Bloch vectors would be in different superpositions (see Fig. \ref{BP12}).  We have also plotted the chessboard diagrams for probability density distribution in position space for $50$ steps of SDC walk with different coins in Fig. \ref{Fig7}.

   \begin{figure*}[!htbp]
  	  \renewcommand*\thesubfigure{\arabic{subfigure}}
   	  \centering
  	  {\begin{tabular}[b]{ccc}
  	  \subfloat[$\theta=0$]{\label{GP0}\includegraphics[width=0.3\textwidth]{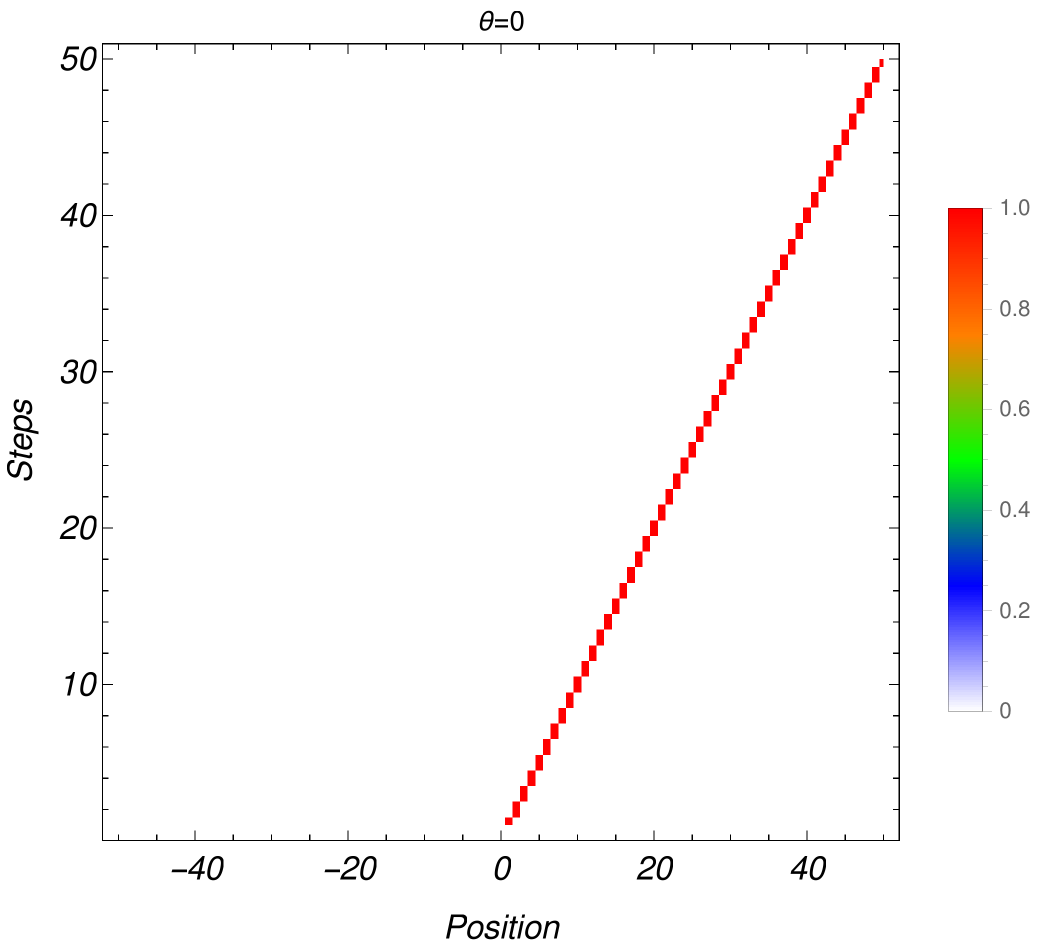}}
  	  \quad 
  	  \subfloat[$\theta=\pi/2$]{\label{GP2}\includegraphics[width=0.3\textwidth]{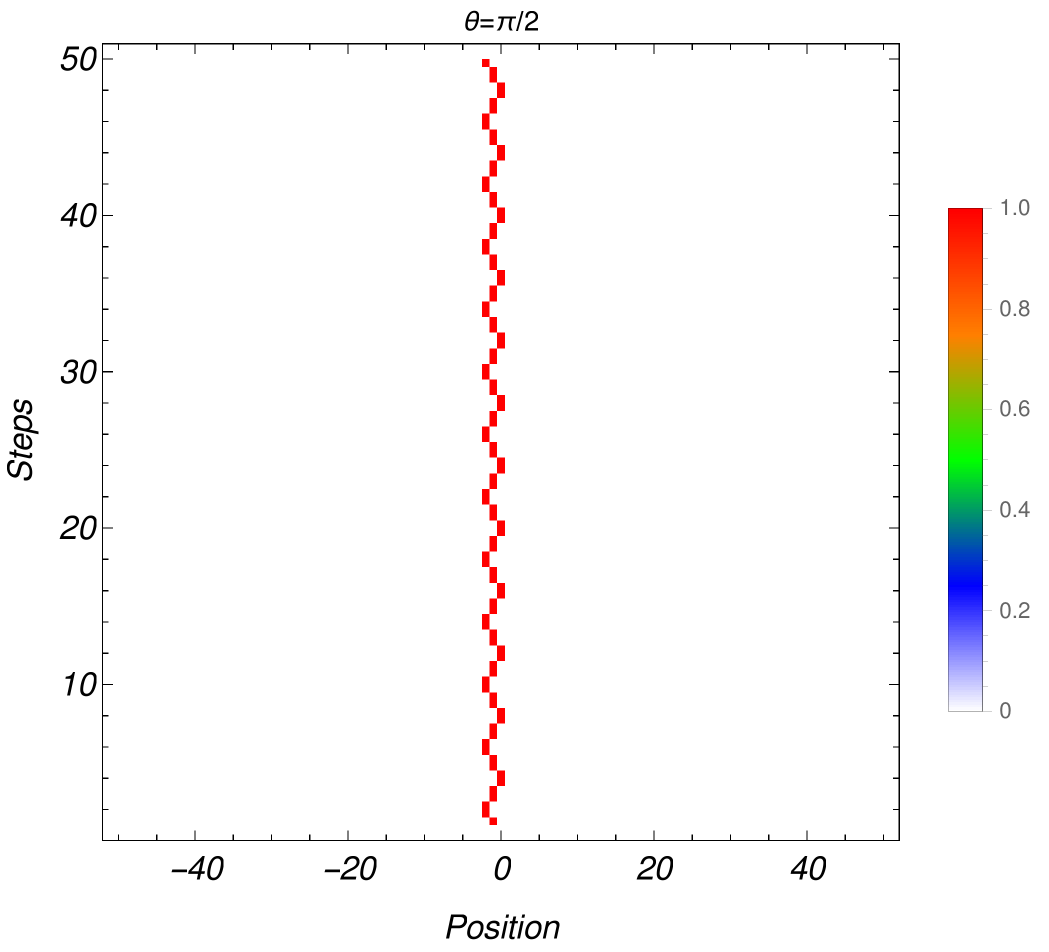}} 	  \quad 
  	  \subfloat[$\theta=\pi/4$]{\label{GP4}\includegraphics[width=0.3\textwidth]{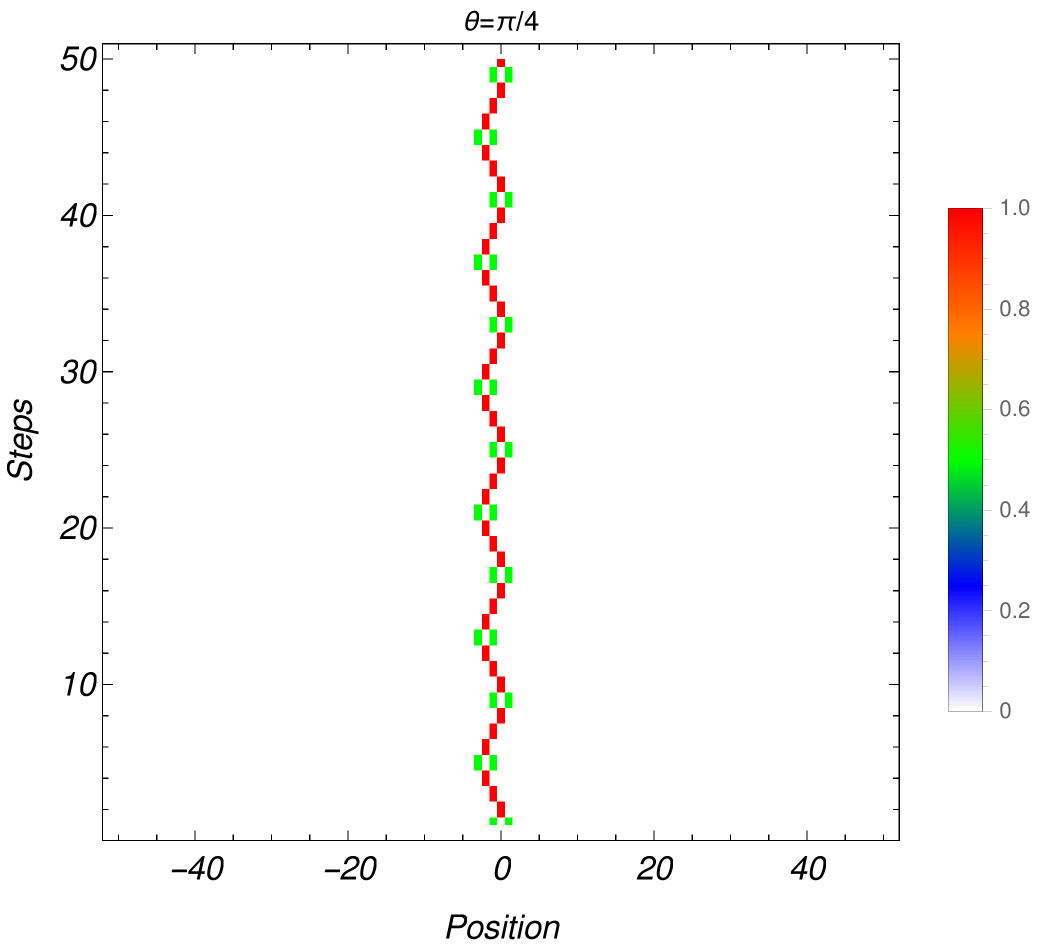}} 	
      \\  	
  	  \subfloat[$\theta=\pi/12$]{\label{GP12}\includegraphics[width=0.3\textwidth]{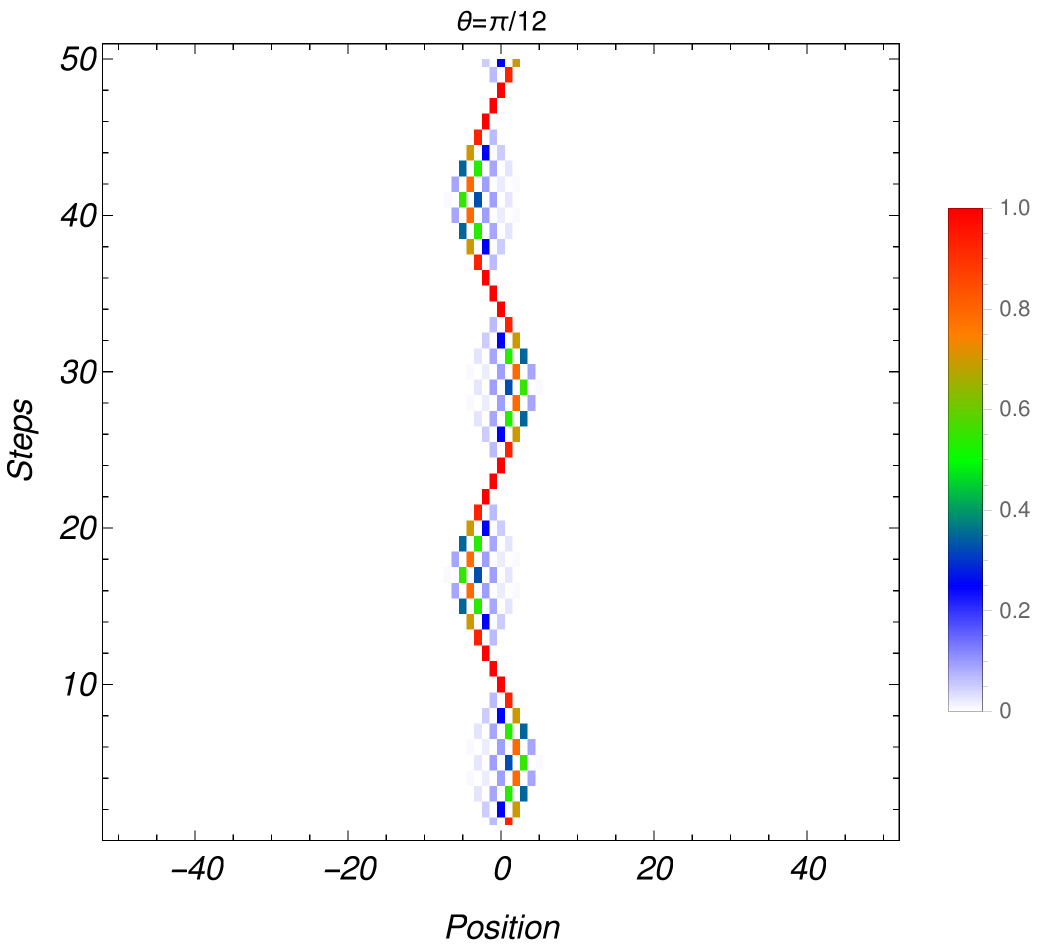}} \quad 
  	  \subfloat[$\theta=3.59\pi/5$]{\label{G15P12}\includegraphics[width=0.3\textwidth]{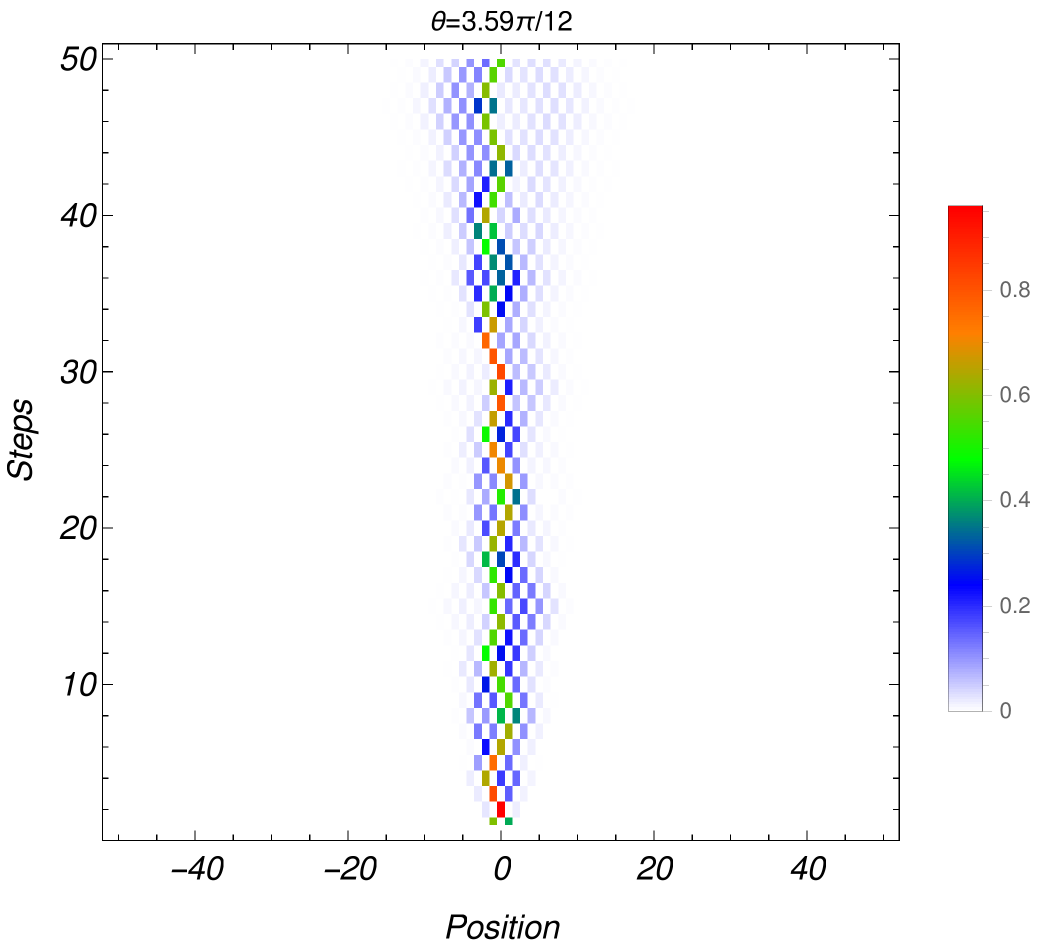}} 
  	  \quad 
  	  \subfloat[$\theta=2\pi/5$]{\label{G2P5}\includegraphics[width=0.3\textwidth]{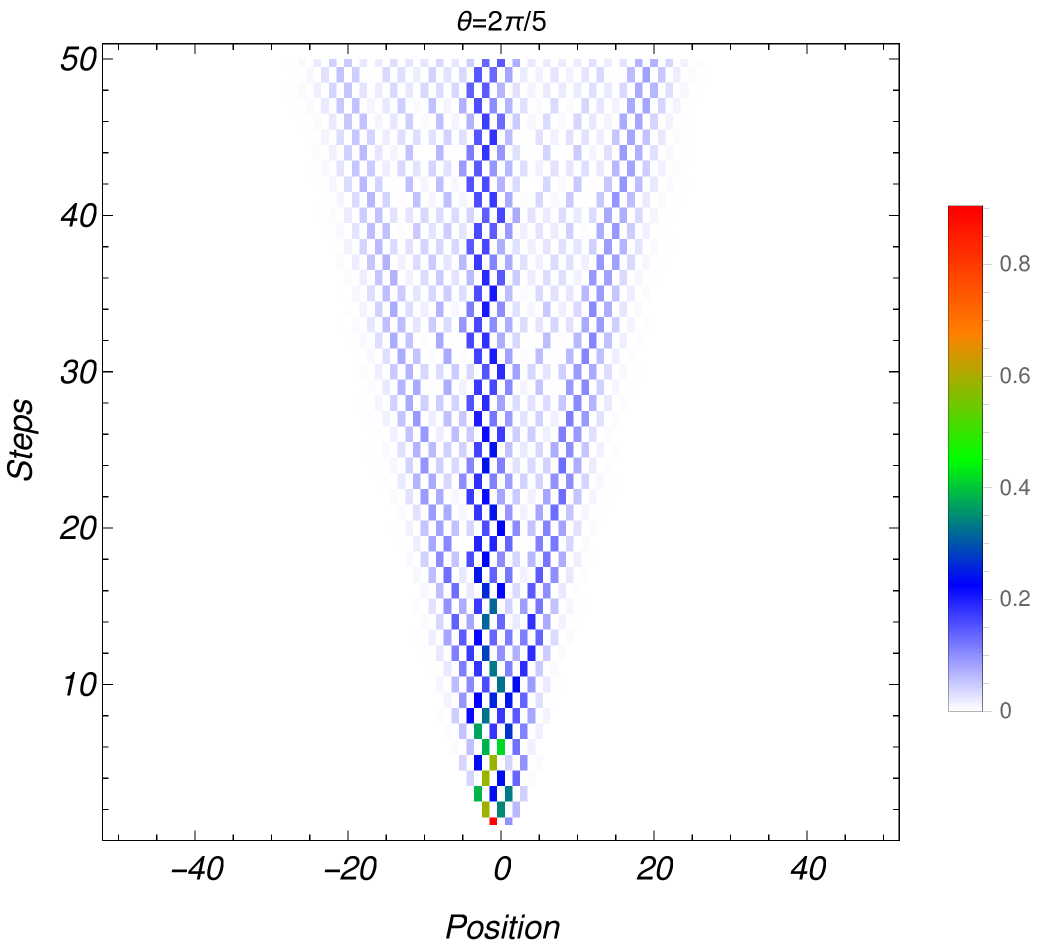}} 
  	  \\ 
  	  \subfloat[$\theta=\pi/5$]{ \label{GP5}\includegraphics[width=0.3\textwidth]{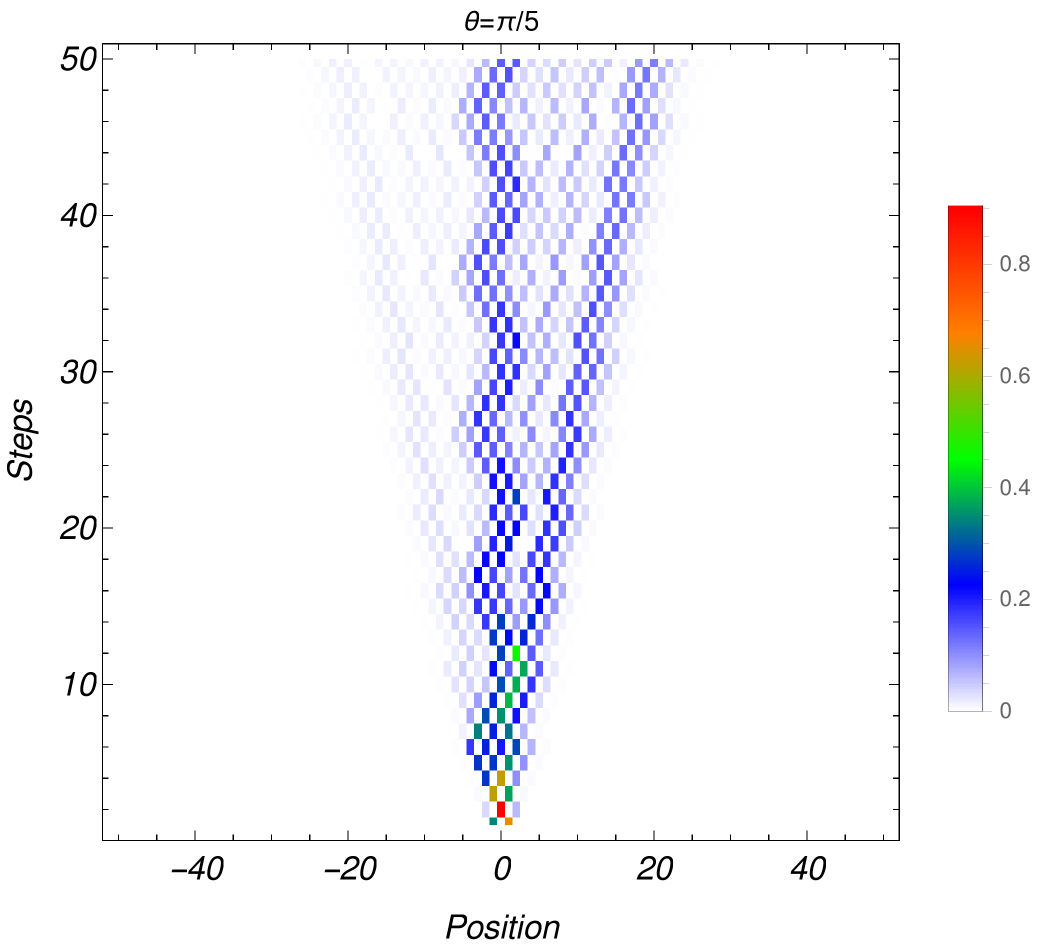}} 	  \quad 
  	  \subfloat[$\theta=\pi/3$]{  \label{GP3}\includegraphics[width=0.3\textwidth]{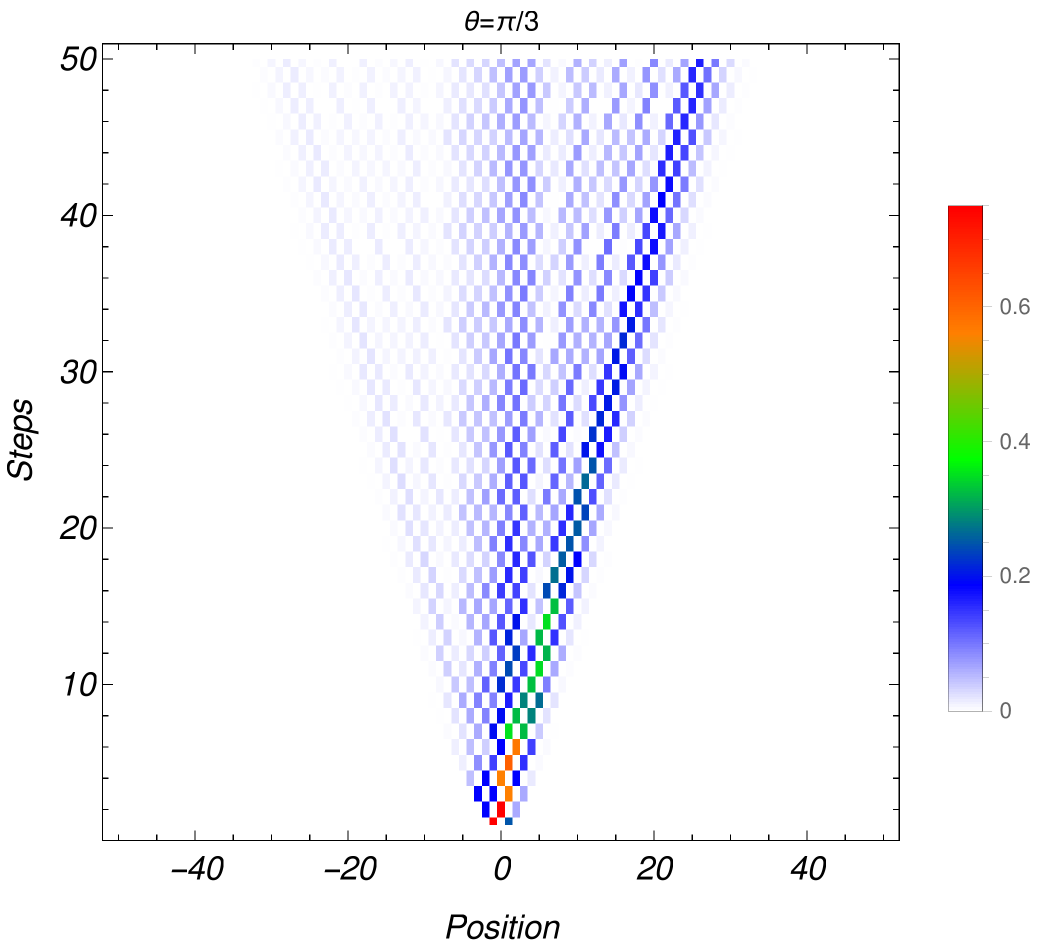}}
  	  \end{tabular}} 	
  	  \caption{Chessboard diagrams for probability density distribution in subsequent steps of $T = 0..50$.}
  	  \label{Fig7}
  	\end{figure*} 
\end{samepage}  	  

\end{document}